\documentstyle[referee,psfig]{l-aa}

\def\etal{{\it et\thinspace al.}\ }

\begin{document}

   \thesaurus{06         
              (03.11.1;  
               16.06.1;  
               19.06.1;  
               19.37.1;  
               19.53.1;  
               19.63.1)} 
   \title{Atomic data from the Iron Project}

\subtitle{XLIV. Transition probabilities and line ratios for Fe~VI
with fluorescent excitation in planetary nebulae}                                                     
   \author{Guo Xin Chen and Anil K. Pradhan}
   \offprints{Guo Xin Chen}
   \institute{ Department of Astronomy, The Ohio State University,
Columbus, OH 43210, USA \\
Internet: chen@astronomy.ohio-state.edu}

   \maketitle

   \begin{abstract}
Relativistic atomic structure calculations for
electric dipole ($E1$), electric quadrupole ($E2$) and 
magnetic dipole ($M1$) transition probabilities  
among the first 80 fine-structure levels of Fe~VI, dominated by
configurations $3d^3, 3d^24s$, and $3d^24p$,
are carried out using the Breit-Pauli version of the code SUPERSTRUCTURE. 
Experimental energies are used to improve the accuracy of  
these transition probabilities. Employing the 80-level
collision-radiative (CR) model with these dipole and forbidden
transition probabilities, and Iron Project R-matrix collisional data, we present
a number of [Fe~VI] line ratios applicable to spectral diagnostics of 
photoionized H~II regions. It is shown that continuum fluorescent excitation
needs to be considered in CR models in order to interpret the
observed line ratios of optical [Fe~VI] lines in planetary nebulae
NGC~6741, IC~351, and NGC~7662. The analysis
leads to parametrization of line ratios as function of, and as
constraints on, the electron
density and temperature, as well as the effective radiation temperature
of the central source and a geometrical dilution factor. The spectral
diagnostics may also help ascertain observational uncertainties. The method may
be generally applicable to other objects with intensive background
radiation fields, such as novae and active galactic nuclei. 
The extensive new Iron Project radiative and collisional calculations
enable a consistent analysis of many line ratios for the complex iron ions.
\footnote{The complete tables of transition probabilities are available
in electronic form at the CDS via anonymous ftp 130.79.128.5}

   \end{abstract}

%

\section {Introduction}

Recently a wealth of optical and UV spectra in Fe VI have been observed
from gaseous nebulae and hot H II regions in general, and from hot
white dwarfs (Jordan \etal 1995). Emission lines from transitions among 
the first 19 fine-structure levels dominated by the ground 
configuration $3d^3$ also appear in nova V 1016 Cygni (Mammano and 
Ciatti, 1975) and RR Telescoppii 
(McKenna \etal 1997). Optical spectra of [Fe VI] have been
observed for many transitions in planetary Nebulae such
as NGC 6741 (Hyung and Aller 1997a), NGC 7662 (Hyung and Aller 1997b), 
IC 351 (Feibelman \etal 1996), and others. It is therefore
interesting to simulate these spectra using accurate atomic data from the
Iron Project (Hummer \etal 1993) obtained using $ab~initio$ calculations. 

 Given the state-of-the-art observational accuracy of astrophysical work,
it is necessary to use accurate atomic radiative and electron
impact excitation (EIE) data in order to calculate accurately the
intensities ratios of prominent density and temperature sensitive lines
of [Fe~VI]. In the collisional-radiative (hereafter CR) model,
all of the level contributions are coupled together and need to be considered.
Furthermore, we have recently shown (Chen and Pradhan 2000; hereafter
CP00) that in
addition to EIE and spontaneous radiative decay, level populations in Fe~VI
are significantly affected by fluorescent excitation (hereafter FLE) via
a radiation backgound, typically a UV continuum. From an atomic physics
point of view this requires additional physical mechanisms to be
included in the atomic model in order to
correctly predict the line intensities. In fact, it was
demonstrated in CP00 that the observed [Fe~VI] optical line ratios in the high
excitation PNe NGC~6741 can not be interpreted without taking account of
the FLE mechanism (as proposed by Lucy 1995; see also Bautista, 
Peng, and Pradhan 1996), which. 
depends on strong
dipole allowed excitations from the ground, or low-lying levels, 
and cascades into the upper levels of observed transitions.

 In order to implement the FLE mechanism in the CR model therefore, one
needs both the forbidden and the dipole allowed transition probabilities,
in addition to the EIE rate coefficients.
Accurate EIE collison strengths and rate coefficients of Fe VI
have recently been computed under the IP by the Ohio-State group using the 
$R$-matrix method. The excitation rate coefficients differ considerably from 
previous works (IP.XXXVII, Chen and Pradhan 1999b; hereafter CP99b).
 The present work has a two-fold aim: (i) to compute transition
proabilities Fe~VI, and (ii) to use a CR model with FLE to compute
line intensity ratios as possible temperature and density diagnostics.
It is further shown that, following CP00, the radiation
temperature of the central source and the distance of the emission
region may be determined if the FLE mechanism is operative. The extended
CR model takes account of the two competing excitation mechanisms, due to
electron excitation (EIE), and photon excitations (FLE).

 A large number of line ratios of [Fe~VI] are examined. 
Density and temperature diagnostics line ratios are determined and
applied to interpret observational data from
a sample of planetary nebulae. The spectral diagnostics developed herein
should have general applications to various other astrophysical sources.

\section{Atomic calculations}

 The EIE calculations are described in detail in CP99a,b, and compared
with previous works. Below, we briefly summarise the qualitative aspects
of those results. The next subsection discusses in detail the present
calculations for the forbidden and allowed A-values for Fe~VI.

\subsection{Electron Impact Excitation of Fe~VI}

With the exception of two calculations two decades ago, there were no
other calculations until the recent work reported in CP99a,b.
It is difficult to consider the relativistic
effects together with the electron correlation effects in this complex
atomic system, and the coupled channel calculations necessary for such
studies are very computer intensive.
The two previous sources for the excitation rates of Fe VI
are the non-relativistic close coupling (CC)
calculations by Garstang \etal (1978) and the distorted-wave (DW)
calculations by Nussbaumer and Storey (1978).
Although the Garstang \etal (1978) calculations were in the CC
approximation,
they used a very small basis set and did not obtain the resonance
structures;
their results are given only for the averaged values.
The Nussbaumer and Storey (1978) calculations were in the DW
approximations that does not enable a treatment of resonances.
Therefore neither set of calculations included resonances or the
coupling effects
due to higher configurations. Owing primarily to these factors we find
that the
earlier excitation rates of Nussbaumer and Storey are lower by up to 
factors of three or more
when compared to the new Fe~VI rates presented in CP99b.

\subsection {Radiative transition probabilities}

The target expansions in the present work are based on
the 34-term wave function expansion for Fe~VI developed by Bautista (1996) 
using the SUPERSTRUCTURE program in the non-relativistic 
calculations for photoionization cross sections of Fe~V.
The SUPERSTRUCTURE calculations for Fe~VI were extended to include relativistic
fine structure using the Breit-Pauli Hamiltonian (Eissner \etal 1974;
Eissner 1998).
The designations for the 80 levels (34 LS terms)
dominated by the configurations
$3d^3$, $3d^24s$ and $3d^24p$
and their observed energies (Sugar and Corliss, 1985), 
are shown
in Table~1. These observed energies were used in the Hamiltonian
diagonalization to obtain the R-matrix surface amplitudes in stage STGH
(Berrington \etal 1995).
This table also provides the key to the level indices
for transitions in tabulating dipole-allowed and forbidden transition
probabilities and 
the Maxwellian-averaged
collision strengths from CP99a,b. Examining the new Breit-Pauli SUPERSTRUCTURE 
calculations we deduce that
the computed energy values for levels 7 and 13; levels 11 and 12
in Table~1 of CP99a should be reversed given the level designations, respectively.
An indication of the
accuracy of the target eigenfunctions may be obtained from
the calculated energy levels in Table~1 of CP99a, and from the computed
length and velocity oscillator strengths for some of the dipole
fine structure transitions given in their Table~2. The agreement between the
length and velocity oscillator strengths is generally about 10\%, an acceptable
level of accuracy for a complex iron ion.

\subsubsection{Dipole allowed fine-structure transitions}

The weighted oscillator strength $gf$ or the Einstein $A$-coefficient for a
dipole allowed fine-structure transition is proportional to the
generalised line strength (Seaton 1987) defined, in either length
form or velocity form, by the equations

\begin{equation}
S_L=\mid <\Psi_j\mid\sum^{N+1}_{k=1}z_k\mid\Psi_i>\mid ^2
\end{equation}

and

\begin{equation}
S_V=\omega^{-2}\mid\left<\Psi_j\mid\sum^{N+1}_{k=1}
\frac{\partial}{\partial z_k}\mid\Psi_i\right >\mid ^2
\end{equation}

where $\omega$ is the incident photon energy in Ry, and $\Psi_i$ and $\Psi_j$
are the wave functions representing the initial and final states, respectively.

Using the transition energy, $E_{ij}$, between the initial and final states,
$g_if_{ij}$ and $A_{ji} $for this transition can be obtained from S as

\begin{equation}
g_if_{ij}=\frac{E_{ij}}3S
\end{equation}

and

\begin{equation}
A_{ji}(a.u.)=\frac 12\alpha ^3\frac{g_i}{g_j}E_{ij}^2f_{ij}
=2.6774\times 10^9(E_j-E_i)^3S^{E1}_{ij}/g_j
\end{equation}

where $\alpha=1/137.036$ is the fine structure constant in a.u.,
and $g_i$, $g_j$ are the
statistical weights of the initial and final states, respectively. In terms
of c.g.s unit of time,

\begin{equation}
A_{ji}(s^{-1})=\frac {A_{ji}(a.u.)}{\tau _0}
\end{equation}
where $\tau _0=2.4191^{-17}s$ is the atomic unit of time.

We can use experimental transition energy $E_{ij}^{exp}$ to obtain refined
$g_if^e_{ij}$ and $A^e_{ji}$ values through
\begin{equation}
g_if^e_{ij}=g_if_{ij}\frac{E_{ij}^{exp}}{E_{ij}^{cal}}
\end{equation}

\begin{equation}
A^e_{ji}=A_{ji}(\frac{E_{ij}^{exp}}{E_{ij}^{cal}})^3
\end{equation}

 Computed $gf_L$ and $gf_V$ values, in both the length
and the velocity formulations,  
for 867 $E1$ (dipole allowed and intercombination) transitions within the first 80 fine structure levels are tabulated
in Table~2 (a partial table is given in the text; the complete Table~2
is available electronically from the CDS library).  Transition probabilities
$A_L$ are also given in the length formulation, which is generally 
more accurate than the velocity formulation in the present calculations.
Experimental level energies are used to improve the accuracy of the
calculated $gf$ and A-values.
All of these $E1$ transition probabilities of
Fe~VI were incorporated in the calculation of line ratios when
accounting for the FLE effect by the UV continuum radiation field
(details below).

\subsubsection{Forbidden electric quadrupole ($E2$) and magnetic dipole ($M1$)
transitions}

The Breit-Pauli mode of the SUPERSTRUCTURE code was also used to 
calculate the $E2$ and the $M1$
transitions in Fe VI. The configuration expansion was adapted from that 
used to optimise the lowest 34 LS terms by Bautista (1996). The
spectroscopic configurations,
the correlation configurations and the scaling
parameters $\lambda _{nl}$ for the Thomas-Fermi-Dirac-Amaldi type
potential of orbital $nl$ are listed in Table~3 and Table~4 of CP99a.
Much effort was devoted to choosing the correlation configurations
to optimise the target wavefunctions, within the constraint of
computational constraints associated with large memory requirements for
many of the $3p$ open shell configurations.
The primary criteria in this
selection are the level of agreement with the observed values for (a) the
level energies and fine structure splittings within the lowest LS terms, 
and (b) the f-values for a number of the low lying dipole allowed
transitions.
Another practical criterion is that the calculated
$A$-values should be relatively stable with minor changes in scaling
parameters.

Like the procedure used in the calculation of the dipole allowed and
intercombination $E1$  $gf$-values, 
the experimental level energies are also used
to improve the accuracy of  the computed $E2$ and $M1$ transition 
probabilities $A^{E2}$ and $A^{M1}$,
given as

\begin{equation}
g_jA^{E2}_{ji}=2.6733\times 10^3(E_j-E_i)^5S^{E2}(i,j) (s^{-1})
\end{equation}

and

\begin{equation}
g_jA^{M1}_{ji}=3.5644\times 10^4(E_j-E_i)^3S^{M1}(i,j) (s^{-1})
\end{equation}

The computed $A^{E2}$ and $A^{M1}$ for all 130 transitions among the
first 19 levels are given in Table~3. The results calculated by
Garstang \etal (1978) and by Nussbaumer and Storey (1978) are also
given for comparison, where available. For 70 transitions in Table~3
the $A^{E2}$ are much smaller than the $A^{M1}$, by up to several orders
of magnitude for some transitions. While for the other 60
transitions, $A^{E2}$ are greater than the $A^{M1}$.
The computed $A^{E2}$ and $A^{M1}$ for all the other 1101 transitions within
the first 80 levels are given in Table~4 (a partial table is given in
the text; the complete Table~4 is available electronically from
the CDS library). 
There are no other calculations
in literature for these transitions for comparisons.
There are many
cases where one of the two transition probabilities is negligible, usually
the $A^{E2}$. But the case of Fe~VI is somewhat different from that of
Fe~III (Nahar and Pradhan 1996), where the
$A^{E2}$ are greater than $A^{M1}$  for nearly half the total number of
transitions, especially those with large excitation energies.

\section{The 80-level collision radiative model with fluorescence}

A CR model with 80 levels for Fe VI is employed to calculate level
populations, $N_i$, relative to the ground level. The emitted flux
per ion for transition $ j\rightarrow i$,
or the emissivity $\epsilon_{ji}$ (ergs cm$^{-3}$s$^{-1}$), is given by,

\begin{equation}
\epsilon_{ji}=N_jA_{ji}h\nu _{ij}
\end{equation}

The line intensities ratios, for example, $\epsilon _{ji}/\epsilon _{kl}$
for transitions j-i and k-l are then calculated.
A computationally efficient code to set up CR matrix in CR model is
employed to solve the coupled linear equations for all levels involved
(Cai and Pradhan 1993).

We can write the rate coupled equations of statistical equilibrium
in CR matrix form:
\begin{equation}
{\bf C=NeQ+A}
\end{equation}
where $Ne$ is the electron density, and
\begin{equation}
C_{ii}=0\\
C_{ij}=q_{ij}Ne ~~~(j>i)\\
C_{ij}=q_{ij}Ne+A_{ij} ~~~(j<i)\\
\end{equation}
In the above model, we have assumed the optically thin case. FLE by continua
pumping and other pumping mechanisms are not considered in this mode.
Also, the line emission photons from the tramsitions within the ions
are assumed to escape directly without absorption (Case A).

In a thermal continuum radiation field of optically thick
gaseous nebulae, the ion can be excited by photon
pumping or de-excited by induced emission.
With the FLE  mechanism for excitation or de-excitation in the CR model,
Eqs. [10] and [11] should be replaced by
\begin{equation}
C_{ij}=q_{ij}Ne+J_{ij}B_{ij} ~~~(j>i)\\
C_{ij}=q_{ij}Ne+A_{ij}+J_{ij}B_{ij} ~~~(j<i)\\
\end{equation}
where $J_{ij}$ is the mean intensity of the continuum at the frequency
for transition $i\rightarrow j$. $B_{ij}$ is the Einstein absorption
coefficient or induced emission coefficients. For a blackbody radiation
with effective temperature $T_{eff}$, $J_{ij}$ can be expressed as

\begin{equation}
J_{ij}=WF_{\nu}=W\frac{8\pi h\nu ^3}{c^2}\frac 1{e^{h\nu/kT_{eff}}-1}
\end{equation}

where $\pi F_{\nu}$ is the monochromatic flux at the photosphere
with an effective temperture $T_{eff}$.
$W=\frac 14(\frac Rr)^2=1.27\times 10^{-16}(\frac
{R_{\ast}/R_{\odot}}{r/pc})^2$
is the geometrical dilution factor, R and r
being the radius of the photosphere and the distance between the star
and the nebula, respectively. More accurate specific luminosity
density function versus frequency, or the mean intensity,
may be employed in ohter applications, e.g. in the synchrotron continuum
pumping in the Crab nebula (Davidson \& Fesen 1985; Lucy 1995).

With the notations used and explanations given above, the equation
of statistical equlibrium for the kth level has the form

\begin{equation}
\sum_{j\neq k}(N_jC_{jk}-N_kC_{kj})=0
\end{equation}

The attenuation effect in continuum intensity $J_{ij}$ has been neglected
in the above rate equations, i.e. there are no optical depths along the
line of observation in the nebula to the continuum radiation source.
This approximation may be responsible for part of the difference between
the calculated and observed line ratios as shown in Table~6 below.
With this approximation, the rate coefficients $C_{ij}$ are independent of
level population $N_k$; the rate equations are therefore linear and
can be solved directly.

\section{[Fe~VI] line ratios: temperature and density diagnostics}

 Some of the salient features of the spectral diagnostics with FLE are
discussed in CP00. It is shown that the line ratios can be parametrized
as a function of {$T_e,N_e,T_{eff},W(r)$}. For a given subset of ($T_e,N_e$)
a line ratio may describe the locus of the subsets of ($T_{eff},W(r)$),
which then defines (constrains) a contour of possible parameters. CP00
present 3-dimensional plots of the line ratio vs. ($T_{eff},W(r)$), for
given ($T_e,N_e$). While it is clear that the $T_{eff}$ or the
$W(r)$ can not be determined uniquely and independently, it was found
that the observed value of the line ratio cuts across the surface
(double-valued function in $T_{eff},W(r)$), along the contour of likely
values that lie within. The variation of the intensity ratio 
with effective temperature and the distance of the emitting region
may constrain these two macrospopic quantities, in addition to the
determination of the local electron temperature and density.

 The spectral diagnostics so developed is applied to the analysis of
[Fe~VI] lines from planetary nebulae as described below.

\subsection{Planetary Nebulae}

 The central stars of planetary nebulae correspond to high stellar
radiation temperatures (e.g. Harman and Seaton 1966), of the order of
10$^5$ K. Resonant absorption was first discussed by
Seaton (1968), who pointed out the efficacy of this mechanism in line
formation of [O~III], in addition to electron scattering and
recombination, estimated the oxygen abundance taking this into account.
 We might expect, a priori, that if the
atomic structure of the emitting ion is subject to FLE then the PNe might
be good candidates for radiative fluorescence studies in general, as
shown in CP00.

In recent years, Hyung and Aller in particular have made a number
of extensive spectral studies of PNe, and in nearly all of those
 [Fe~VI] optical emission lines have been detected (Table~5). 
Physical conditions in some of the PNe's are listed in Table~5. Observed
line ratios are used to develop the temperature-density diagnostics
for [Fe~VI]. An earlier study of [Fe~VI] line ratios was carried out by
Nussbaumer and Storey (1978) who calculated a number of line
emissivities relative to the $\lambda$ 5146 $\AA$ line. Owing to new
atomic collisional and radiative data our line ratios differ
significantly from the earlier work for many lines. Also, Nussbaumer and
Storey (1978) did not take the FLE mechanism into account. On
examination of the observed [Fe~VI] lines in several PNe, we noted
that the $\lambda$ 5146 $\AA$ line was mis-identified and 
assigned to [O~I] in the PNe labeled a,c,d in Table~5 (in a private
communication we confirmed the new
identification with Prof. Lawrence Aller, who
noted that [Fe~VI] was the more likely source, 
particularly in NGC 6741  which is a high excitation object). 

 A comprehensive study of most of the possible line ratios was carried out as 
functions of
$T_e$, $N_e$, and with and without FLE, at various radiation temperatures
T$_{eff}$ and dilution factors W($R_{\ast}/r)$.
In Table~6 we present line ratios for lines which frequently
appear in various kinds of PNe's, with different physical conditions,
with respect to the line $\lambda$ 5146$\AA$ (as in
Nussbaumer and Storey 1978). The partial Table~6 given in the text
contains only those line ratios for which observed values are available.
The complete Table~6 (available electronically from the
CDS library), gives a number of other line ratios computes in a similar manner
relative to the $\lambda$ 5146$\AA$.

Line ratios are calculated with different
dilution factors within the CR model in order to evaluate the influence
of FLE under different conditions.
The first four entries are:  no FLE,
FLE with $W=5\times 10^{-16}, 10^{-13}, 10^{-10}$ respectively.
The Nussbaumer and Storey (1978) values are given as the fifth set of
entries for comparison. Also, observational values are give in these entries
(under ``Obs") for various planetary nebulae, wherever available.

\subsection{NGC 6741}

 Observations of this high excitation nebula by \cite{all85} and \cite{hyu97} show
several optical [Fe~VI] lines in the spectrum from the multiplet $3d^3 \ (^4F - ^4P)$
at 5177, 5278, 5335, 5425, 5427, 5485, 5631 and 5677 $\AA$ and from
the $(^4F - ^2G)$ at 4973 and 5146 $\AA$ for NGC 6741. The basic observational
parameters, in particular the inner and the outer radii needed to
estimate the distance from the central star and the dilution factor, 
are described in these works, and their diagnostic diagrams based on the
spectra of a number of
ions give $T_e = 12500K, N_e = 6300 cm^{-3}$, and a stellar $T_{eff} = 
140,000 K$.
As the ionization potentials of Fe~V and Fe~VI are 75.5 eV and 100 eV
respectively, compared to that of He~II at 54.4 eV, 
Fe~VI emission should stem from the fully ionized $He^{2+}$ zone, and
within the inner radius, i.e. r(Fe~VI) $\leq r_{in}$. With these
parameters we obtain the dilution factor to be W = 10$^{-14}$; 
the dominant [Fe~VI] emission region could be up to a factor of 3 closer to
the star, with W up to 10$^{-13}$, without large variations in the
results obtained.

 Figs. 1 and 2 show all the line ratios for NGC 6741
(with respect to the 5146 $\AA$ line), where observational
values are available (Hyung and Aller 1997a).
[Fe~VI] line ratios are presented as a function of several
parameters, in particular with and without FLE. 
In all cases the FLE = 0 curve fails to correlate with the
observed line ratios, and shows no dependence on N$_e$ (an unphysical
result), whereas with FLE we obtain
a consistent $N_e \approx$ 1000-2000 cm$^{-3}$, 
suitable for the high ionization [Fe~VI] zone. The derived
N$_e$ is somewhat lower than the N$_e$ range 2000 - 6300 cm$^{-3}$ 
obtained from several ionic spectra (including [O~II] and [S~II]) 
by \cite{hyu97}.
The total observational uncertainties cited by Hyung and Aller (1997a)
are 17.6\%, 19.5\%, 38.9\%, 15.6\%, 23.2\%, 25.5\%, 10.2\%, 14.5\%, and 36.5\%
for 4973, 5177, 5278, 5335, 5425, 5427, 5485, 5631 and 5677 $\AA$
(with respect to the 5146 $\AA$ line),
respectively (Hyung and Aller 1997a).
However, an indication of the overall uncertainties
may be obtained from the 1st line ratio, 4973/5146,
which is independent of both T$_e$ and N$_e$ since both lines have the 
same upper level, and which therefore 
depends only on the ratio of the A-values and energy separations; the observed value of 1.048
agrees closely with the theoretical value of 0.964.
Whereas the combined observational and theoretical uncertainties for 
any one line ratio can be significant, 
most measured line ratios (except three line ratios 5278/5146, 5427/5146 and 5677/5146 which will
be analyzed in the next paragraph) yield a remarkably consistent N$_e$([Fe~VI]) and
substantiate the spectral model with FLE.
 
While the electron density of the [Fe VI] in NGC 6741 is determined to be 
$\approx$ 1000-2000 cm$^{-3}$ from most of the observational line ratios as demonstrated
above, we note from Fig. 2 that 
three line ratios 5278/5146, 5427/5146 and 5677/5146 deviate from 
this N$_e$ considerably.
It is interesting to estimate the possible errors in these three observational
line ratios from our theoretical method and model. 
Several pairs of line ratios are shown in Table~7 that have a common 
upper level. These line ratios depend only on the ratio of the
A-values and energy differences and
 are
independent of the detailed physical conditions in PN. They can be 
used as important means to determine possible, systematic errors in 
the observed line intensties as described below.

 From the line ratio 
4973/5146, independent of PN physical conditions as mentioned above, 
and the other five line ratios in Table~7,
we conclude that the intensity of the reference line 5146 $\AA$ should be 
very accurate.
It is very unlikely that the error for each pair of these line ratios 
are the same
and show the same tendency. An error estimate of 6.8\% for this line 
given by Hyung and Aller (1997a) is consistent with our justification. As such, the
observed intensities of lines $\lambda\lambda$ 4973, 5177, 5335, 
5425, 5485, and 5631 $\AA$ should be of high
accuracy (within 20\%). This conclusion is also supported by the good agreement between the two
other theoretical and observational line ratios 5335/5485 and 5425/5632 shown in
Table~7.

Based on these arguments, we find
that the observed (`Hamilton') line intensity of 0.087   (Hyung and Aller 1997a)
(relative to the uniform flux of I(H$\beta$)=100)
for 5278 $\AA$ should be reduced by about 40\% from a
comparison of the line ratio 5278/5425 in Table~7.
Similarly, the reported line intensity of 0.118 for the line 5427 $\AA$ 
should be reduced by about 70\% or more, and the value 0.122 for the
5677 $\AA$ should be
increased by 20\% or so from the
comparison of the line ratio 5427/5677 in Table~7. If our justifications for 
the errors in the intensities of these three lines are correct, the 
corresponding three line ratios shown in Fig. 2 will also yield the same 
and consistent N$_e$ as do the
other line ratios, particularly those in Fig. 1. It is interesting to note that the uncertainties given by 
Hyung and Aller (1997a) are also large (as inferred above), although the 
line 5427 $\AA$ could have a much
higher uncertainty (intensity larger or lower).

\subsection{IC 351}
 
The physical conditions of IC 351, especially the effective temperature T$_{eff}$
and the distance of PN emission region to the central white dwarf (WD), 
are highly uncertain.
We first apply our method, as developed in CP00, to determine the 
appropriate T$_{eff}$ and the
dilution factor W(r). T$_{eff}$ is thereby determined to be 
80,000$\pm$10,000 K. This is considerably different
from T$_{eff}$=58,100K cited by \cite{fei96}. It is interesting to note that
the T$_{eff}$ determined by our spetral method with FLE agrees with the
He~II Zanstra temperature, which is 85,000K according to \cite{pre83}. 
As pointed out
by \cite{pre83}, different methods (Zanstra method, color temperature and
energy balance method, etc.) used to determine the effective temperature of 
PN remain discrepant;
but the He~II Zanstra method is applicable to optically thick PN's. With
the T$_{eff}$ as above, we obtain the dilution factor W(r)
to be 10$^{-13}$ -- 10$^{-14}$.

After determining T$_{eff}$ and W(r), the same method as used in NGC 6741 
is applied to determine the electron density N$_e$ of the [Fe~VI] 
emission nebula in IC 351,
and possible errors in the observed line intensities. 
There are 7 observational line
ratios for IC 351 as shown in Figs. 3 and 4; but we calculated the same 
8 line ratios theoretically as for NGC 6741.
The only reported uncertainty for a line ratio given by \cite{fei96} is 
36.6\% for the pair 5335/5146 (Fig. 3).
When we look into the comparison of line ratios for IC 351 from Table~7
(pairs of line ratios with common upper level), we find that
the agreement of 4973/5146 is good ($\approx$ 10\%) between calculation and observation.
This establishes that the intensities of both the 4973 and 5146 $\AA$
lines should be highly
accurate ($\approx$ 10\%). The diffierence is 26\% for 5427/5677 
(Fig. 4), 92\% for 5335/5485 (Fig. 3), and 175\% for 5278/5425 (Fig. 4). 
From these differences, and Figs. 3 and 4, we could estimate the 
possible errors 
in some of the observed lines: line 5485 $\AA$
should be reduced by 60\%; line 5336 $\AA$ increased by 20\%; 
line 5278 $\AA$ reduced by
175\%. The intensity of the line 5425 $\AA$ is accurate. From these 
4 line ratios,
N$_e$ is determined to be $\approx$ 1,000 cm$^{-3}$. 
Using this N$_e$ and Fig. 4, 
and the comparison of the line ratio 5427/5677 in Table~7, the possible
errors in the intensities of the
other 3 observed lines can be deduced as follows: line 5427 $\AA$
should be reduced by 80\%; line 5677 $\AA$ reduced by 40\%; and line 5177 $\AA$
increased by 20\%. In summary, the above error analysis is based on 
Table~7 and the theoretically computed line ratios reported in this work
(Figs. 3 and 4).

\subsection{NGC 7662}

Finally we apply our spectral diagnostics, and the same procedures used 
above, to NGC 7662. In this PN, the effective temperature T$_{eff}$ and 
emission region distance
to the central star seem to have been determined within low uncertainties. 
Hence, we
adopt here T$_{eff}$=105,000 K and W=10$^{-14}$ (\cite{hyu97b}) in our present
calculation and the results are shown in Figs. 5 and 6. However, the observational uncertainties
in line intensities
are even larger than those in \cite{fei96} as shown from both the rms uncertainties given
in \cite{hyu97b} and our detailed spectral analysis by using the method 
developed by Chen and Pradhan (2000). The more complex problem is
that our reference line 5146 $\AA$ has very large observational
uncertainty in NGC 7662, in contrast to NGC 6741 or IC 351. 
Therefore we need to determine the uncertainty of this reference line 
first, before
the general spectral diagonosis of NGC 7662 including N$_e$ can proceed.

That the uncertainty in observational intensity 
of line 5146 $\AA$ is large can be established from the comparison of 
the physical-condition-independent line ratio 4973/5146 shown in Table~7 for NGC 7662.
We also have some supplemental evidence to support this conclusion. 
First, after some
analysis, we find the observed intensity of line 5425 $\AA$ to be very accurate;
at most it is too high by 5-10\%. This is also consistent with 
the rms uncertainty given by \cite{hyu97b}, which is remarkbly low at 3.9\%.
Second, we could establish that the observed intensities for lines
5335 and 5177 $\AA$ are also very accurate (within 10\%) by using our 
procedures as described above.
It is very likely that only the 5335 $\AA$ line intensity needs to be 
increased by 5\%, and the line
5177 $\AA$ intensity decreased by 5\%. To further confirm our justification, 
we plot a new line
ratio 5177/5335 in Fig. 7, which the leads to a reasonable determination
of N$_e$. In all the above arguments, we have a consistent spectral analysis 
from the line ratios of
5335/5146, 5177/5146 and 5425/5146 as shown in Figs. 5 and 6, as well as the 
line ratio 5177/5335 in Fig. 7.
 As such, we conclude that: (1) the observed intensity of
line 5146 should be inreased by 80\%. (the rms given in \cite{hyu97b} 
is 21.7\%);
(2) the electron density of the [Fe~VI] emission region is 
N$_e\approx$3,000 cm$^{-3}$.
For this PN, we can then deduce the possible observational errors in the other 
lines comparing the line ratios 5335/5485 and 5425/5631 shown in Table~7.
We find that the observed intensity of the line 5485 $\AA$ should be 
decreased by 35\%; and that of line 5631 $\AA$ increased by 60\%. 
Finally, to make the observational line ratio
5677/5146 consistent with the physical conditions including N$_e$ 
determined above by
all the other line ratios, the observed intensity of the line 5677 $\AA$ 
should be increased by 110\% (the rms uncertainty given by \cite{hyu97b} 
is 20.3\%).

\section{Conclusion and discussion}

 An extensive calculation of fine structure transition probabilities of
Fe~VI is presented for the allowed $E1$ and the forbidden $E2$, $M1$ 
transitions.  An indication of the uncertainties in the computed
gf-values is given in the plot of length $gf_L$ vs. the velocity $gf_V$
for 867 $E1$ transitions computed in this work (Fig. 8). 
It shows an agreement at about 10\% level for most of the
transitions, with no more than about 5\% of the transitions lying
outside that range even for gf-values less than 10$^{-4}$.

Combined with previously calculated data for electron impact excitation,
a 80-level CR spectral model for line ratios diagnostics is
used to predict the effect of collisional and fluorescent excitation
(FLE) in planetary nebulae. An illustrative and limited analysis of 
line ratios is carried out as an example of the use of the atomic data
and the model proposed herein.
Some of the diagnostics procedures developed earlier by Chen and Pradhan 
(2000, CP00) are
employed to analyse observed line intensties from three planetary
nebulae: NGC 6741, IC 351 and NGC 7662. The detailed model and analysis
yields a consistent set of diagnostics, for example, for the electron
density and effective temperature of the source. It shows
that fluorescence effects should be included in CR
models of these objects. 
The method may be used to determine the physical conditions 
in PN, especially when it is difficult to do so by other
methods. For example, we have determined, within certain error limits, 
both the effective
temperature and the emission region distance (via a dilution factor) 
for IC 351.
 By combining the line ratios that are independent of the physical conditions 
of PN (like cases in Table~5), and our method, we are able to estimate
possible errors in the observed intensities,
both individually and in term of consistency among the set of observed
lines. It is expected that the method and procedures described in this
paper would be generally applicable to spectral diagnostis of other 
radiative plasma sources, such as novae and AGN.

 Finally, some possible uncertainties in our model and
procedure, as employed in the present calculations, are as follows. 
1) Static conditions are
assumed in the CR model, independent of photoionization eqilibrium in
the [Fe~VI] regions of PN's. Because [Fe~VI] region is almost the highest
ionization state in PN's considered here, this assumption should not carry
a large uncertainty. 2) We have assumed inherently in our procedures that
there are nearly constant N$e$ and T$e$ in [Fe~VI] regions. This assumption
is not entirely true because both N$e$ and T$e$ are a function of the
distance from the central star to the [Fe~VI] emission region. However,
according to 1), if photoionization equilibrium prevails then 
there should not  be
large variations in N$e$ and T$e$.
3)  The radiation field should in principle simulate the ionizing
stellar radiation. A further refinement of the model proposed herein
would be (a) to include a radiation field with proper allowance for the
Helium and Hydrogen opacities in various ionization and excitation
steps, and 
(b) in addition to the radiation flux from the central star, 
resonance fluorescence from H~I and He~II Ly$\alpha$ should be
considered in the model.
 However as noted earlier, Fe~VI is likely to be in the fully
ionized He~III zone, and therefore not greatly susceptible to the
effects.
4) Even though the most advanced R-matrix codes are employed in
generating atomic data, the atomic data still have some uncertainties,
estimated at about 10--20\%. 

 All data tables are available electronically from the CDS, 
or via ftp from the authors at: chen@astronomy.ohio-state.edu.

\begin{acknowledgements}
This work was supported by a grant (AST-9870089) 
from the U.S. National Science Foundation and by NASA grant NAG5-8423.
The calculations were carried out on the massively
parallel Cray T3E and the vector processor Cray T94 at
the Ohio Supercomputer Center in Columbus, Ohio.

\end{acknowledgements}

\def\amp{{\it Adv. At. Molec. Phys.}\ }
\def\apj{{\it Astrophys. J.}\ }
\def\apjs{{\it Astrophys. J. Suppl. Ser.}\ }
\def\apjl{{\it Astrophys. J. (Letters)}\ }
\def\aj{{\it Astron. J.}\ }
\def\aa{{\it Astron. Astrophys.}\ }
\def\aasup{{\it Astron. Astrophys. Suppl.}\ }
\def\adndt{{\it At. Data Nucl. Data Tables}\ }
\def\cpc{{\it Comput. Phys. Commun.}\ }
\def\jqsrt{{\it J. Quant. Spectrosc. Radiat. Transfer}\ }
\def\jpb{{\it Journal Of Physics B}\ }
\def\pasp{{\it Pub. Astron. Soc. Pacific}\ }
\def\mn{{\it Mon. Not. R. astr. Soc.}\ }
\def\pra{{\it Physical Review A}\ }
\def\prl{{\it Physical Review Letters}\ }
\def\zpds{{\it Z. Phys. D Suppl.}\ }

 \ \\
 \ \\
 \ \\
 \ \\
 \ \\
 \ \\
 \ \\
 \ \\
 \ \\
\begin{table}
\caption{The 80 fine structure levels 
corresponding to the 34 $LS$ terms included in the calculations
and their observed energies (Ry)
in Fe~VI (Sugar and Corliss, 1985).}
\begin{center}
\begin{tabular} {rllrlcrllrl}
\hline
 i & Term & & 2J & Energy & \hspace*{0.3in} & i & Term & & 2J & Energy \\
\hline
1&3d$^3$&$^4$F&3&0.0&&41&3d$^2(^3$F)4p&$^4$F$^o$&5&3.101443 \\
2&&&5&0.004659&&42&&&7&3.110750 \\
3&&&7&0.010829&&43&&&9&3.120492 \\
4&&&9&0.018231&&44&3d$^2(^3$F)4p&$^2$F$^o$&5&3.121742 \\
5&3d$^3$&$^4$P&1&0.170756&&45&&&7&3.131189 \\
6&&&3&0.172612&&46&3d$^2(^3$F)4p&$^4$D$^o$&1&3.131290 \\
7&&&5&0.178707&&47&&&3&3.127568 \\
8&3d$^3$&$^2$G&7&0.187870&&48&&&5&3.137250 \\
9&&&9&0.194237&&49&&&7&3.147723 \\
10&3d$^3$&$^2$P&1&0.241445&&50&3d$^2(^3$F)4p&$^2$D$^o$&3&3.140706 \\
11&&&3&0.238888&&51&&&5&3.152138 \\
12&3d$^3$&$^2{\rm D}2$&3&0.260877&&52&3d$^2(^3$F)4p&$^2$G$^o$&7&3.179977 \\
13&&&5&0.259568&&53&&&9&3.189597 \\
14&3d$^3$&$^2$H&9&0.261755&&54&3d$^2(^3$P)4p&$^2$S$^o$&1&3.205891 \\
15&&&11&0.266116&&55&3d$^2(^3$P)4p&$^4$S$^o$&3&3.240986 \\
16&3d$^3$&$^2$F&5&0.424684&&56&3d$^2(^1$D)4p&$^2$P$^o$&1&3.272354 \\
17&&&7&0.421163&&57&&&3&3.260106 \\
18&3d$^3$&$^2$D$1$&3&0.656558&&58&3d$^2(^1$D)4p&$^2$F$^o$&5&3.265386 \\
19&&&5&0.653448&&59&&&7&3.279505 \\
20&3d$^2(^3$F)4s&$^4$F&3&2.386075&&60&3d$^2(^3$P)4p&$^4$D$^o$&1&3.275057\\
21&&&5&2.390877&&61&&&3&3.278569 \\
22&&&7&2.397871&&62&&&5&3.287005 \\
23&&&9&2.406823&&63&&&7&3.301248 \\
24&3d$^2(^3$F)4s&$^2$F&5&2.452586&
&64&3d$^2(^1$D)4p&$^2$D$^o$&3&3.297495 \\
25&&&7&2.466551&&65&&&5&3.304281 \\
26&3d$^2(^1$D)4s&$^2$D&3&2.562646&
&66&3d$^2(^3$P)4p&$^4$P$^o$&1&3.316518 \\
27&&&5&2.559763&&67&&&3&3.320593 \\
28&3d$^2(^3$P)4s&$^4$P&1&2.565008&&68&&&5&3.330627 \\
29&&&3&2.570092&&69&3d$^2(^1$G)4p&$^2$G$^o$&7&3.326827 \\
30&&&5&2.578448&&70&&&9&3.328555 \\
31&3d$^2(^3$P)4s&$^2$P&1&2.623713&&71&3d$^2(^3$P)4p&$^2$D$^o$&3&
3.376592 \\
32&&&3&2.630266&&72&&&5&3.376970 \\
33&3d$^2(^1$G)4s&$^2$G&7&2.663908&&73&3d$^2(^1$G)4p&$^2$H$^o$&9&
3.390785 \\
34&&&9&2.663752&&74&&&11&3.405461 \\
35&3d$^2(^1$S)4s&$^2$S&1&3.065870&&75&3d$^2(^3$P)4p&$^2$P$^o$&1&
3.408944 \\
36&3d$^2(^3$F)4p&$^4$G$^o$&5&3.082420&&76&&&3&3.412018 \\
37&&&7&3.093542&&77&3d$^2(^1$G)4p&$^2$F$^o$&5&3.454410 \\
38&&&9&3.106829&&78&&&7&3.444151 \\
39&&&11&3.123191&&79&3d$^2(^1$S)4p&$^2$P$^o$&1&3.719860 \\
40&3d$^2(^3$F)4p&$^4$F$^o$&3&3.094115&&80&&&3&3.739745 \\
\hline
\end{tabular}
\end{center}
\end{table}
 \ \\

\newpage
\begin{table}
\caption{Partial Table 2 (complete table available electronically from
CDS). Weighted dipole allowed $E1$ oscillator strengths
$gf_L$, $gf_V$ in the length and velocity formulations, and the Einstein A-coefficients $A_L$ in the length formulation.}
\begin{tabular} {lrlll|lrlll}
\hline
i&j&$gf_L$&$gf_V$&$A_L$&i&j&$gf_L$&$gf_V$&$A_L$\\
\hline
36& 1&9.34e-02&9.68e-02&1.19e+09&
80& 2&9.64e-06&9.10e-06&2.70e+05\\
40& 1&4.64e-01&4.68e-01&8.92e+09&
36& 3&2.91e-04&2.88e-04&3.68e+06\\
41& 1&9.82e-02&9.85e-02&1.26e+09&
37& 3&2.14e-04&2.67e-04&2.04e+06\\
44& 1&2.43e-03&2.50e-03&3.17e+07&
38& 3&1.75e-01&1.80e-01&1.34e+09\\
46& 1&2.57e-01&2.55e-01&1.01e+10&
41& 3&1.05e-01&1.05e-01&1.34e+09\\
47& 1&2.88e-02&2.89e-02&5.66e+08&
42& 3&9.02e-01&9.05e-01&8.71e+09\\
48& 1&8.98e-04&8.83e-04&1.18e+07&
43& 3&8.20e-02&8.14e-02&6.37e+08\\
50& 1&5.48e-02&5.40e-02&1.09e+09&
44& 3&8.66e-02&8.58e-02&1.12e+09\\
51& 1&2.56e-03&2.50e-03&3.40e+07&
45& 3&2.86e-02&2.91e-02&2.80e+08\\
54& 1&1.82e-06&1.86e-06&7.49e+04&
48& 3&5.06e-01&5.01e-01&6.62e+09\\
55& 1&1.90e-08&1.54e-08&4.01e+02&
49& 3&3.99e-02&3.90e-02&3.95e+08\\
56& 1&5.62e-03&5.89e-03&2.42e+08&
51& 3&8.14e-02&8.18e-02&1.07e+09\\
57& 1&1.22e-04&1.29e-04&2.61e+06&
52& 3&6.25e-04&5.87e-04&6.30e+06\\
58& 1&1.12e-04&1.09e-04&1.59e+06&
53& 3&1.30e-03&1.17e-03&1.06e+07\\
60& 1&6.74e-02&7.11e-02&2.90e+09&
58& 3&1.12e-02&1.16e-02&1.58e+08\\
61& 1&2.92e-02&3.06e-02&6.29e+08&
59& 3&6.04e-03&6.33e-03&6.48e+07\\
62& 1&2.05e-03&2.14e-03&2.97e+07&
62& 3&1.71e-01&1.77e-01&2.45e+09\\
64& 1&4.64e-04&4.80e-04&1.01e+07&
63& 3&2.73e-02&2.80e-02&2.97e+08\\
65& 1&5.92e-05&5.99e-05&8.65e+05&
65& 3&3.28e-03&3.35e-03&4.77e+07\\
66& 1&2.69e-05&2.67e-05&1.19e+06&
68& 3&1.32e-04&1.28e-04&1.95e+06\\
67& 1&2.30e-05&2.23e-05&5.09e+05&
69& 3&9.30e-05&8.67e-05&1.03e+06\\
68& 1&9.37e-07&8.67e-07&1.39e+04&
70& 3&1.45e-06&1.80e-07&1.28e+04\\
71& 1&1.64e-04&1.37e-04&3.74e+06&
72& 3&2.46e-04&1.83e-04&3.74e+06\\
72& 1&2.17e-05&1.86e-05&3.32e+05&
73& 3&2.64e-06&2.50e-06&2.42e+04\\
75& 1&2.27e-05&1.54e-05&1.06e+06&
77& 3&1.38e-05&1.58e-05&2.19e+05\\
76& 1&1.93e-06&1.28e-06&4.50e+04&
78& 3&7.28e-07&8.26e-07&8.62e+03\\
77& 1&1.51e-05&1.87e-05&2.42e+05&
37& 4&8.83e-04&8.95e-04&8.38e+06\\
79& 1&1.30e-05&1.26e-05&7.22e+05&
38& 4&1.59e-03&1.52e-03&1.22e+07\\
80& 1&2.25e-06&2.15e-06&6.33e+04&
39& 4&1.86e-01&1.90e-01&1.20e+09\\
36& 2&2.29e-03&2.47e-03&2.90e+07&
42& 4&7.97e-02&8.01e-02&7.66e+08\\
37& 2&1.36e-01&1.41e-01&1.31e+09&
43& 4&1.29e+0&1.29e+0&1.00e+10\\
40& 2&8.40e-02&8.49e-02&1.61e+09&
45& 4&3.29e-01&3.28e-01&3.20e+09\\
41& 2&6.24e-01&6.27e-01&8.01e+09&
49& 4&6.18e-01&6.10e-01&6.07e+09\\
42& 2&1.20e-01&1.19e-01&1.16e+09&
52& 4&8.53e-05&8.03e-05&8.57e+05\\
44& 2&5.67e-03&5.86e-03&7.37e+07&
53& 4&4.35e-03&4.02e-03&3.52e+07\\
45& 2&3.71e-03&3.76e-03&3.64e+07&
59& 4&5.10e-02&5.22e-02&5.45e+08\\
47& 2&3.04e-01&3.01e-01&5.96e+09&
63& 4&2.30e-01&2.36e-01&2.49e+09\\
48& 2&6.17e-02&6.10e-02&8.10e+08&
69& 4&3.62e-04&3.81e-04&3.97e+06\\
49& 2&3.61e-04&3.41e-04&3.58e+06&
70& 4&5.33e-05&4.02e-05&4.70e+05\\
50& 2&1.34e-01&1.34e-01&2.65e+09&
73& 4&2.09e-07&8.11e-08&1.91e+03\\
51& 2&2.82e-02&2.79e-02&3.74e+08&
74& 4&1.18e-05&1.07e-05&9.04e+04\\
52& 2&5.53e-04&4.96e-04&5.60e+06&
78& 4&1.12e-04&1.21e-04&1.32e+06\\
55& 2&4.58e-07&4.10e-07&9.63e+03&
40& 5&8.03e-04&7.85e-04&1.38e+07\\
57& 2&5.81e-04&6.07e-04&1.24e+07&
46& 5&1.14e-01&1.09e-01&4.03e+09\\
58& 2&2.26e-03&2.43e-03&3.21e+07&
47& 5&8.80e-02&8.41e-02&1.54e+09\\
59& 2&2.86e-04&2.88e-04&3.08e+06&
50& 5&2.53e-02&2.40e-02&4.48e+08\\
61& 2&1.17e-01&1.22e-01&2.51e+09&
54& 5&1.56e-03&1.57e-03&5.78e+07\\
62& 2&3.70e-02&3.84e-02&5.33e+08&
55& 5&1.66e-01&1.67e-01&3.15e+09\\
63& 2&1.42e-03&1.47e-03&1.55e+07&
56& 5&1.88e-03&1.90e-03&7.28e+07\\
64& 2&8.48e-04&8.65e-04&1.85e+07&
57& 5&5.31e-03&5.48e-03&1.02e+08\\
65& 2&9.55e-04&9.84e-04&1.39e+07&
60& 5&2.36e-02&2.32e-02&9.15e+08\\
67& 2&1.22e-04&1.21e-04&2.70e+06&
61& 5&2.81e-02&2.77e-02&5.46e+08\\
68& 2&1.99e-05&1.81e-05&2.94e+05&
64& 5&1.97e-03&1.97e-03&3.88e+07\\
69& 2&3.84e-06&2.14e-06&4.26e+04&
66& 5&1.95e-02&1.95e-02&7.75e+08\\
71& 2&7.81e-05&5.55e-05&1.78e+06&
67& 5&8.78e-02&8.79e-02&1.75e+09\\
72& 2&1.82e-04&1.48e-04&2.77e+06&
71& 5&2.20e-04&2.10e-04&4.55e+06\\
76& 2&1.63e-05&1.15e-05&3.81e+05&
75& 5&3.78e-04&3.79e-04&1.59e+07\\
77& 2&1.80e-06&1.96e-06&2.87e+04&
76& 5&4.32e-05&2.90e-05&9.11e+05\\
78& 2&1.07e-05&1.33e-05&1.27e+05&
79& 5&2.11e-05&1.94e-05&1.07e+06\\
\hline
\end{tabular}
\end{table}

\newpage
\begin{table}
\caption{Comparison of the electric quadrupole ($E2$) and 
magnetic dipole ($M1$) transition probabilities 
between the first 19 levels of Fe VI in $s^{-1}$
among the present calculation by SUPERSTRUCTURE, calculation by
Garstang $et~al.$ (1978) and calculation by Nussbaumer and Storey
(1978).}
\begin{tabular} {llllllll}
\hline
\multicolumn{2}{l}{transition}&
\multicolumn{2}{c}{Present}&
\multicolumn{2}{c}{Garstang $et~al.$}&
\multicolumn{2}{c}{NS}\\
i&j&E2&M1&E2&M1&E2&M1\\
\hline
 2& 1&5.13e-11&5.76e-3&0.0&5.7e-3&4.97-11&5.74-3\\
 5& 1&6.04e-2&2.01e-4&8.3e-2&8.0e-5&5.97-2&3.31-4\\
 6& 1&1.27e-2&3.40e-3&1.7e-2&1.2e-3&1.26e-2&4.05e-3\\
 7& 1&
7.15e-4&2.15e-4&
1.0e-3&9.0e-5&7.04e-4&2.66e-4\\
 8& 1&1.90e-5&0.0&1.4e-5&0.0&1.66-5&\\
10& 1&1.99e-3&1.54e-3&7.0e-3&7.3e-4&1.54e-3&1.99e-3\\
11& 1&
6.88e-4&3.70e-1&
2.8e-3&1.19e-1&5.40e-4&3.56e-1\\
12& 1&
3.84e-4&3.77e-1&
6.6e-4&4.01e-1&2.67e-4&3.86e-1\\
13& 1&
5.71e-7&4.97e-2&
2.8e-6&3.36e-2&9.53e-7&4.34e-2\\
16& 1&4.62e-3&1.97e-1&&&4.53e-3&2.23e-1\\
17& 1&6.60e-4&0.0&&&6.55e-4&\\
18& 1&3.69e-3&8.60e-2&&&4.14e-3&1.26e-1\\
19& 1&6.35e-4&4.97e-3&&&6.30e-4&9.44e-3\\
 3& 2&2.03e-10&1.34e-2&0.0&1.3e-2&1.99e-10&1.34e-2\\
 4& 2&6.46e-10&0.0&&&&\\
 5& 2&3.47e-2&0.0&4.85e-2&0.0&3.42e-2&\\
 6& 2&3.35e-2&1.97e-3&4.59e-2&6e-4&3.32e-2&1.78e-3\\
 7& 2&
5.69e-3&9.87e-4&
7.9e-3&4.2e-4&5.63e-3&1.36e-3\\
 8& 2&1.63e-6&2.04e-1&1.6e-5&1.73e-1&1.74e-6&2.44e-1\\
 9& 2&6.00e-6&0.0&2e-6&0.0&4.71e-6&\\
10& 2&3.19e-3&0.0&1.5e-3&0.0&2.75e-3&\\
11& 2&
3.39e-3&5.98e-1&
6.4e-3&1.84e-1&2.78e-3&5.75e-1\\
12& 2&
7.96e-4&7.82e-1&
1.2e-3&7.1e-1&5.37e-4&7.30e-1\\
13& 2&
2.95e-5&1.44e-1&
2.1e-7&9.5e-2&2.79e-5&1.39e-1\\
14& 2&2.59e-5&0.0&2.5e-5&0.0&2.70e-5&\\
16& 2&9.77e-4&2.73e-2&&&1.09e-3&3.08e-2\\
17& 2&1.90e-3&8.39e-2&&&1.70e-3&1.01e-1\\
18& 2&2.77e-6&1.81e-1&&&1.21e-5&2.51e-1\\
19& 2&2.05e-3&1.53e-2&&&2.11e-3&2.39e-2\\
 4& 3&3.66e-10&1.44e-2&0.0&1.4e-2&3.59e-10&1.45e-2\\
 6& 3&3.86e-2&0.0&5.4e-2&0.0&3.84e-2&\\
 7& 3&
2.12e-2&2.39e-3&
2.94e-2&9e-4&2.11e-2&2.63e-3\\
 8& 3&1.83e-5&2.19e-1&1.2e-5&1.85e-1&2.02e-5&2.61e-1\\
 9& 3&2.02e-6&2.20e-1&3.1e-5&1.86e-1&1.96e-6&2.51e-1\\
11& 3&
4.95e-3&0.0&
6.1e-3&0.0&4.14e-3&\\
12& 3&
2.18e-3&0.0&
5.3e-4&0.0&1.71e-3&\\
13& 3&
3.82e-4&1.12e+0&
5e-5&7.29e-1&3.41e-4&1.07e+0\\
14& 3&6.41e-5&2.20e-3&1.6e-5&4.3e-3&7.12e-5&4.12e-3\\
15& 3&1.83e-5&0.0&5.0e-5&0.0&1.94e-5&\\
16& 3&9.55e-4&3.57e-2&&&1.06e-3&3.78e-2\\
17& 3&4.11e-4&1.49e-2&&&5.97e-4&1.70e-2\\
18& 3&1.30e-2&0.0&&&1.57e-2&\\
19& 3&4.04e-3&1.64e-1&&&4.94e-3&2.45e-1\\
 7& 4&
5.28e-2&0.0&
7.3e-2&0.0&5.23e-2&\\
 8& 4&4.06e-6&1.26e-2&8.2e-6&1.1e-2&4.29e-6&1.34e-2\\
 9& 4&7.96e-5&5.39e-1&7.8e-5&4.55e-1&8.62e-5&6.24e-1\\
13& 4&
7.74e-4&0.0&
5.1e-4&0.0&6.12e-4&\\
14& 4&5.55e-6&3.35e-3&2.3e-5&7.8e-3&4.63e-6&6.86e-3\\
15& 4&1.56e-4&6.73e-4&1.9e-4&5.7e-4&1.68e-4&1.01e-3\\
16& 4&1.60e-4&0.0&&&2.04e-4&\\
17& 4&4.27e-3&2.17e-1&&&5.01e-3&2.56e-1\\
19& 4&5.55e-2&0.0&&&6.41e-2&\\
 6& 5&6.56e-13&1.86e-4&0.0&1.85e-4&6.64e-13&1.87e-4\\
 7& 5&
5.71e-9&0.0&
&&&\\
10& 5&0.0&4.42e-1&0.0&3.3e-1&&3.76e-1\\
11& 5&
6.09e-7&1.10e-1&
2e-7&8.1e-2&4.88e-7&9.29e-2\\
12& 5&
3.67e-7&2.08e-2&
1.26e-5&1.3e-3&1.71e-7&1.52e-2\\
13& 5&
3.22e-7&0.0&
5.7e-6&0.0&2.84e-7&\\
16& 5&8.51e-4&0.0&&&5.80e-4&\\
\hline
\end{tabular}
\end{table}

\begin{table}
\centerline {Table 3. {\it continued}}
\begin{tabular} {llllllll}
\hline
\multicolumn{2}{l}{transition}&
\multicolumn{2}{c}{Present}&
\multicolumn{2}{c}{Garstang $et~al.$}&
\multicolumn{2}{c}{NS}\\
i&j&E2&M1&E2&M1&E2&M1\\
\hline
18& 5&3.83e-2&1.26e-1&&&2.99e-2&1.37e-1\\
19& 5&9.70e-3&0.0&&&7.16e-3&\\
 7& 6&
2.08e-9&4.70e-3&
0.0&4.6e-3&2.09e-9&4.73e-3\\
10& 6&1.46e-7&9.14e-6&1.5e-6&3.6e-5&2.21e-7&1.76e-5\\
11& 6&
3.61e-9&2.56e-1&
9.5e-6&1.7e-1&1.69e-8&2.13e-1\\
12& 6&
1.00e-5&1.02e-2&
5.4e-6&7.4e-4&8.43e-6&6.67e-3\\
13& 6&
7.01e-7&2.19e-3&
5.4e-5&1.8e-3&5.06e-7&2.27e-3\\
16& 6&1.30e-6&5.25e-4&&&2.13e-8&1.04e-3\\
17& 6&3.04e-3&0.0&&&2.15e-3&\\
18& 6&1.05e-1&4.73e-1&&&8.08e-2&5.20e-1\\
19& 6&1.02e-1&2.20e-1&&&7.71e-2&2.41e-1\\
 8& 7&
4.85e-13&1.12e-10&
&&&\\
10& 7&
2.97e-6&0.0&
1.0e-5&0.0&2.45e-6&\\
11& 7&
5.13e-6&1.14e-1&
1.5e-5&1.19e-1&4.17e-6&1.00e-1\\
12& 7&
1.31e-6&2.19e-1&
3.6e-6&7.4e-2&9.72e-7&1.76e-1\\
13& 7&
5.45e-6&7.98e-2&
3.7e-6&4.3e-2&4.33e-6&6.71e-2\\
14& 7&
1.76e-9&0.0&
&&&\\
16& 7&
7.10e-5&2.32e-3&
&&6.11e-5&5.47e-3\\
17& 7&
3.12e-4&2.11e-4&
&&2.69e-4&4.95e-4\\
18& 7&
8.14e-4&1.06e-1&
&&7.99e-4&1.23e-1\\
19& 7&
1.49e-3&1.30e+0&
&&1.18e-3&1.41e+0\\
 9& 8&1.41e-12&4.03e-3&0.0&4e-3&1.71e-13&4.01e-3\\
11& 8&
1.12e-5&0.0&
4.9e-6&0.0&7.85e-6&\\
12& 8&
6.73e-5&0.0&
9.3e-5&0.0&5.13e-5&\\
13& 8&
5.80e-6&2.08e-6&
9.6e-6&9e-6&3.89e-6&1.06e-5\\
14& 8&1.82e-4&8.33e-2&1.9e-4&1.38e-1&1.59e-4&1.24e-1\\
15& 8&5.32e-6&0.0&1.7e-5&0.0&4.28e-6&\\
16& 8&1.47e-1&1.35e-1&&&1.49e-1&1.49e-1\\
17& 8&1.24e-2&2.29e-1&&&1.24e-2&2.53e-1\\
18& 8&1.25e+01&0.0&&&1.18e+1&\\
19& 8&9.45e-1&1.99e-3&&&9.07e-1&2.23e-3\\
13& 9&
5.83e-5&0.0&
5.9e-5&0.0&4.08e-5&\\
14& 9&2.32e-5&1.43e-1&0.0&2.35e-1&2.53e-5&2.11e-1\\
15& 9&1.35e-4&7.84e-2&2.2e-4&1.29e-1&1.13e-4&1.16e-1\\
16& 9&7.32e-5&0.0&&&2.06e-5&\\
17& 9&1.24e-1&1.07e-1&&&1.26e-1&1.17e-1\\
19& 9&1.07e+1&0.0&&&1.00e+1&\\
11&10&
1.32e-12&2.19e-4&
0.0&3.2e-4&1.21e-12&2.30e-4\\
12&10&
3.15e-7&4.08e-2&
4.0e-7&1.18e-2&2.94e-7&3.80e-2\\
13&10&
7.59e-8&0.0&
1e-7&0.0&&\\
16&10&1.63e-2&0.0&&&1.41e-2&\\
18&10&1.54e+0&5.98e-3&&&1.43e+0&2.21e-3\\
19&10&6.28e-1&0.0&&&5.87e-1&\\
12&11&5.15e-8&1.04e-1&3e-7&6.56e-2&5.89e-8&1.02e-1\\
13&11&
8.66e-8&5.96e-2&
8e-7&2.6e-2&1.10e-7&5.56e-2\\
16&11&1.05e-2&6.60e-3&&&1.07e-2&8.08e-3\\
17&11&2.26e-2&0.0&&&2.13e-2&\\
18&11&2.11e+0&2.14e-2&&&2.08e+0&1.08e-2\\
19&11&6.72e-1&2.98e-1&&&6.63e-1&3.47e-1\\
13&12&7.95e-13&2.54e-5&0.0&3.4e-5&6.88e-13&2.70e-5\\
16&12&2.19e-2&5.41e-3&&&2.30e-2&7.71e-3\\
17&12&9.32e-4&0.0&&&4.44e-4&\\
18&12&5.98e-2&9.58e-3&&&2.56e-2&3.81e-3\\
19&12&1.58e+0&2.69e-1&&&1.46e+0&3.09e-1\\
14&13&6.89e-15&0.0&&&&\\
16&13&
7.25e-3&2.75e-2&
&&7.66e-3&3.63e-2\\
17&13&
2.92e-2&1.02e-2&
&&3.10e-2&1.36e-2\\
18&13&
2.73e-1&7.48e-1&
&&2.96e-1&8.64e-1\\
19&13&
5.79e-1&4.77e-3&
&&6.22e-1&4.80e-3\\
15&14&5.22e-11&1.33e-3&0.0&1.3e-3&5.33e-11&1.32e-3\\
16&14&7.02e-2&0.0&&&6.78e-2&\\
17&14&1.20e-3&5.00e-4&&&8.85e-4&8.13e-4\\
\hline
\end{tabular}
\end{table}

\begin{table}
\centerline {Table 3. {\it continued}}
\begin{tabular} {llllllll}
\hline
\multicolumn{2}{l}{transition}&
\multicolumn{2}{c}{Present}&
\multicolumn{2}{c}{Garstang $et~al.$}&
\multicolumn{2}{c}{NS}\\
i&j&E2&M1&E2&M1&E2&M1\\
\hline
19&14&1.19e-1&0.0&&&1.50e-1&\\
17&15&5.18e-2&0.0&&&4.99e-2&\\
17&16&3.12e-12&8.88e-4&&&2.88e-12&8.86e-4\\
18&16&5.14e-1&3.95e-1&&&4.67e-1&3.72e-1\\
19&16&9.09e-2&6.41e-1&&&8.26e-2&6.01e-1\\
18&17&1.05e-1&0.0&&&9.67e-2&\\
19&17&5.24e-1&3.73e-1&&&4.75e-1&3.50e-1\\
19&18&7.24e-11&6.42e-4&&&6.68e-11&6.41e-4\\
\hline
\end{tabular}
\end{table}

\newpage
\begin{table}
\caption{Partial Table 4 (complete table available from CDS). Electric quadrupole ($E2$) and 
magnetic dipole ($M1$) transition probabilities 
between the first 80 levels of Fe VI in $s^{-1}$
in the present calculation by SUPERSTRUCTURE.}
\begin{tabular} {lrll|lrll|lrll|lrll}
\hline
i&j&E2&M1&i&j&E2&M1&i&j&E2&M1&i&j&E2&M1\\
\hline
20& 1&4.12e+4&6.84e-6&
30& 5&9.36e+3&0.0&
26&10&8.54e+3&5.52e-5&
33&14&6.87e+4&2.90e-3\\
21& 1&2.68e+4&2.50e-5&
31& 5&0.0&2.86e-2&
27&10&3.32e+3&0.0&
34&14&1.68e+3&5.30e-3\\
22& 1&2.51e+3&0.0&
32& 5&1.12e+2&6.88e-3&
28&10&0.0&2.53e-4&
22&15&1.19e+2&0.0\\
24& 1&2.63e-1&7.50e-4&
35& 5&0.0&1.97e-4&
29&10&8.11e+3&4.29e-6&
23&15&1.49e+1&9.31e-7\\
25& 1&3.57e-2&0.0&
20& 6&4.29e+3&3.09e-8&
30&10&4.02e+3&0.0&
25&15&5.88e+4&0.0\\
26& 1&4.13e+3&8.11e-4&
21& 6&8.71e+3&1.42e-6&
31&10&0.0&2.34e-4&
33&15&2.31e+3&0.0\\
27& 1&1.48e+2&7.62e-5&
22& 6&9.26e+3&0.0&
32&10&1.07e+4&7.71e-5&
34&15&6.46e+4&2.90e-3\\
28& 1&4.20e+4&1.93e-6&
24& 6&1.64e+2&1.77e-7&
35&10&0.0&3.08e-2&
20&16&7.74e+0&1.87e-3\\
29& 1&4.24e+3&9.91e-4&
25& 6&8.96e+1&0.0&
20&11&2.65e+1&1.39e-4&
21&16&2.15e+0&2.14e-4\\
30& 1&2.52e+2&6.16e-5&
26& 6&1.03e+4&2.65e-4&
21&11&7.38e+1&1.50e-4&
22&16&1.96e+0&3.00e-4\\
31& 1&3.18e+0&8.46e-6&
27& 6&1.35e+4&7.67e-4&
22&11&1.33e+2&0.0&
23&16&1.27e-1&0.0\\
32& 1&8.58e-1&1.28e-8&
28& 6&5.62e+3&2.54e-3&
24&11&1.74e+4&1.13e-4&
24&16&4.36e+3&7.11e-6\\
33& 1&1.58e+0&0.0&
29& 6&7.95e+3&4.01e-4&
25&11&1.47e+3&0.0&
25&16&4.18e+2&2.74e-3\\
35& 1&1.71e+1&4.03e-8&
30& 6&1.09e+4&2.04e-3&
26&11&9.61e+2&3.38e-4&
26&16&5.33e+3&2.20e-4\\
20& 2&3.94e+4&6.70e-5&
31& 6&1.01e+3&5.91e-6&
27&11&1.21e+4&6.87e-4&
27&16&1.29e+3&4.16e-4\\
21& 2&2.23e+4&2.14e-6&
32& 6&2.00e+2&1.28e-2&
28&11&1.44e+2&4.15e-5&
28&16&3.22e+1&0.0\\
22& 2&2.64e+4&5.43e-5&
33& 6&5.07e+0&0.0&
29&11&5.10e+1&1.27e-4&
29&16&4.78e+3&2.44e-4\\
23& 2&1.58e+3&0.0&
35& 6&6.08e+1&5.35e-4&
30&11&1.44e+4&6.58e-5&
30&16&7.93e+2&2.95e-4\\
24& 2&8.11e+1&1.59e-4&
20& 7&3.03e+2&1.62e-7&
31&11&3.62e+4&1.53e-3&
31&16&7.47e+4&0.0\\
25& 2&8.71e+0&2.89e-4&
21& 7&1.82e+3&2.99e-8&
32&11&8.82e+2&4.32e-3&
32&16&1.05e+4&7.40e-6\\
26& 2&1.35e+4&1.68e-3&
22& 7&6.14e+3&1.67e-6&
33&11&7.35e+2&0.0&
33&16&1.99e+4&2.94e-3\\
27& 2&1.46e+3&1.66e-4&
23& 7&1.57e+4&0.0&
35&11&1.35e+4&7.76e-3&
34&16&1.63e+3&0.0\\
28& 2&2.79e+4&0.0&
24& 7&1.35e+0&3.52e-7&
20&12&1.05e+1&1.84e-4&
35&16&1.65e+0&0.0\\
29& 2&1.20e+4&1.59e-3&
25& 7&2.62e+1&1.38e-7&
21&12&2.02e+1&1.89e-4&
20&17&1.57e+0&0.0\\
30& 2&2.16e+3&1.62e-4&
26& 7&1.97e+4&2.31e-3&
22&12&1.15e+1&0.0&
21&17&8.39e+0&4.87e-4\\
31& 2&5.90e+1&0.0&
27& 7&7.67e+3&3.71e-4&
24&12&4.20e+3&1.12e-4&
22&17&4.52e+0&1.60e-4\\
32& 2&1.83e+0&9.06e-6&
28& 7&5.16e+4&0.0&
25&12&9.66e+3&0.0&
23&17&2.32e+0&9.69e-4\\
33& 2&9.02e-1&4.30e-4&
29& 7&1.67e+4&1.33e-3&
26&12&1.58e+4&3.10e-4&
24&17&8.42e+2&4.30e-3\\
34& 2&4.37e-1&0.0&
30& 7&8.72e+3&1.20e-4&
27&12&6.28e+2&2.44e-4&
25&17&5.30e+3&2.47e-6\\
35& 2&1.07e+1&0.0&
31& 7&4.13e+1&0.0&
28&12&2.46e+0&8.28e-4&
26&17&1.34e+3&0.0\\
20& 3&4.85e+3&0.0&
32& 7&7.14e+1&2.29e-2&
29&12&1.98e+4&1.02e-3&
27&17&7.05e+3&1.73e-4\\
21& 3&3.42e+4&7.43e-5&
33& 7&1.15e+0&1.19e-8&
30&12&1.14e+3&1.46e-7&
29&17&2.32e+3&0.0\\
22& 3&3.42e+4&4.87e-7&
34& 7&1.33e+1&0.0&
31&12&8.56e+2&1.40e-3&
30&17&4.42e+3&1.16e-4\\
23& 3&1.92e+4&5.77e-5&
35& 7&1.03e+2&0.0&
32&12&2.18e+4&1.60e-3&
32&17&6.00e+4&0.0\\
24& 3&1.39e+1&7.62e-5&
20& 8&3.60e+1&0.0&
33&12&1.08e+3&0.0&
33&17&2.56e+3&6.63e-3\\
25& 3&1.12e+2&5.86e-5&
21& 8&2.84e+1&6.92e-7&
35&12&1.73e+4&7.24e-3&
34&17&2.11e+4&3.02e-3\\
26& 3&1.85e+4&0.0&
22& 8&3.39e+1&3.79e-6&
20&13&4.21e-2&5.84e-5&
20&18&9.03e-2&6.88e-5\\
27& 3&6.86e+3&1.79e-3&
23& 8&4.37e+0&2.34e-7&
21&13&6.17e-3&7.29e-5&
21&18&1.49e+0&7.92e-5\\
29& 3&1.70e+4&0.0&
24& 8&5.13e+4&1.57e-3&
22&13&2.83e+0&4.27e-4&
22&18&1.79e-1&0.0\\
30& 3&9.20e+3&1.12e-3&
25& 8&5.08e+3&2.61e-3&
23&13&3.95e+1&0.0&
24&18&1.91e+3&5.14e-4\\
32& 3&5.14e+1&0.0&
26& 8&2.18e+4&0.0&
24&13&3.25e+3&4.91e-4&
25&18&3.05e+2&0.0\\
33& 3&3.68e+0&5.18e-4&
27& 8&2.78e+3&6.25e-6&
25&13&1.95e+4&2.33e-4&
26&18&2.74e-1&1.55e-7\\
34& 3&1.39e-1&4.18e-4&
29& 8&2.42e+4&0.0&
26&13&5.49e+3&1.78e-3&
27&18&2.07e+1&1.74e-3\\
21& 4&2.56e+3&0.0&
30& 8&1.98e+3&3.17e-6&
27&13&1.80e+4&1.47e-3&
28&18&1.00e-1&1.11e-6\\
22& 4&2.33e+4&3.83e-5&
32& 8&1.50e+2&0.0&
28&13&1.86e+2&0.0&
29&18&1.03e+1&1.57e-6\\
23& 4&6.63e+4&3.92e-7&
33& 8&2.98e+4&3.66e-9&
29&13&7.66e+3&8.91e-6&
30&18&6.96e+0&1.15e-3\\
24& 4&6.64e+0&0.0&
34& 8&2.61e+3&3.57e-6&
30&13&1.38e+4&2.22e-3&
31&18&1.75e+4&2.77e-6\\
25& 4&1.56e+1&9.41e-4&
21& 9&5.39e+1&0.0&
31&13&6.57e+3&0.0&
32&18&9.36e+3&2.45e-6\\
27& 4&2.07e+4&0.0&
22& 9&1.80e+1&6.01e-6&
32&13&1.20e+4&1.77e-3&
33&18&7.33e+3&0.0\\
30& 4&2.88e+4&0.0&
23& 9&5.95e+1&1.57e-7&
33&13&1.55e+2&2.46e-6&
35&18&4.70e+4&5.07e-8\\
33& 4&7.33e-1&2.94e-5&
24& 9&1.07e+4&0.0&
34&13&2.05e+3&0.0&
20&19&6.10e-2&1.49e-5\\
34& 4&1.21e+1&1.19e-3&
25& 9&4.87e+4&1.31e-3&
35&13&4.51e+4&0.0&
21&19&8.13e-1&5.63e-6\\
20& 5&1.07e+4&1.55e-7&
27& 9&2.20e+4&0.0&
21&14&1.09e+2&0.0&
22&19&2.28e+0&6.04e-5\\
21& 5&4.78e+3&0.0&
30& 9&1.72e+4&0.0&
22&14&2.36e+1&6.08e-7&
23&19&3.46e-2&0.0\\
24& 5&2.55e+1&0.0&
33& 9&8.36e+2&8.76e-5&
23&14&2.01e+0&2.23e-6&
24&19&4.45e+2&1.69e-3\\
26& 5&9.80e+2&2.75e-4&
34& 9&3.35e+4&1.05e-4&
24&14&5.53e+4&0.0&
25&19&2.46e+3&8.39e-4\\
27& 5&8.16e+3&0.0&
20&10&3.97e+1&1.06e-7&
25&14&5.14e+3&7.81e-6&
26&19&3.22e+1&1.91e-3\\
28& 5&0.0&2.17e-5&
21&10&2.69e+1&0.0&
27&14&8.39e+2&0.0&
27&19&6.67e+0&9.49e-6\\
29& 5&2.04e+3&1.39e-3&
24&10&6.36e+3&0.0&
30&14&5.22e+2&0.0&
28&19&8.86e+0&0.0\\
\hline
\end{tabular}
\end{table}

\newpage
\begin{table}
\caption{Physical conditions of gaseous nebulae}
\begin{tabular} {cccccccc}
\hline
Source&$N_e/10^3cm^{-3}$&$T_e/10^3K$&$T_{eff}/K$&r/pc&R/$R_{\odot}$&$W$&Reference\\
\hline
 NGC 6741&6.3&12.5&140&0.0063&0.063&$1.3\times 10^{-14}$&Hyung and
Aller 1997a
; a\\
 NGC 6886&5 - 10&13&150&0.001(0.0345)&0.046&$2.7\times 10^{-13}$&Hyung \etal
1995; b\\
 NGC 6884&10&10&110&0.002(0.020)&0.13&$5.4\times 10^{-13}$&Hyung \etal 1997; c\\
 IC 351&2.5 - 20&13 - 16&58.1&0.05&0.72&$2.6\times 10^{-14}$&Feibelman \etal 1996; d\\
 NGC 2440&5&14.2&180&0.015(0.0425)&0.038&$8.2\times 10^{-16}$&Hyung and
Aller 1998; e\\
 NGC 7662&3 - 17&13&105&0.025(0.035)&0.15&$4.6\times 10^{-15}$&Hyung
and Aller 1997b; f\\
\hline
\end{tabular}
\\
$^a$ Hyung and Aller (1997a);
$^b$ Hyung \etal (1995);
$^c$ Hyung \etal (1997);
$^d$ Feibelman \etal (1996);\\
$^e$ Hyung and Aller (1998);
$^f$ Hyung and Aller (1997b).
\end{table}
\clearpage

\newpage
\begin{table}
\caption{Partial Table 6 (complete table available electronically from
CDS). Line intensity ratios for transitions relative to 
$\lambda$5146 $AA$ (8-3: $^4F_{7/2}$--$^2G_{7/2}$), with T$_{eff}$=150,000
K, and with observed values from planetary
nebulae in the fifth column (the full table 6 contains a number of other line ratios).
The first four entries are for no FLE, and
FLE with $W=5\times 10^{-16}, 10^{-13}, 10^{-10}$, respectively.
Entries in the fifth column are values calculated by
Nussbaumer and Storey (1978).}
\begin{tabular} {cccccccccccc}
\hline
&&\multicolumn{2}{c}{$T_e$=10000}&\multicolumn{2}{c}{$T_e$=20000}
&\multicolumn{3}{c}{$T_e$=12000}
&\multicolumn{3}{c}{$T_e$=16000}\\
Transition&$\lambda(\AA)$&$n_e=10^3$&$10^4$&$10^3$&$10^4$
&$2\times 10^3$&$6\times 10^3$&$10^4$
&$2\times 10^3$&$6\times 10^3$&$10^4$\\
\hline
8-2&4973&9.64-1&9.64-1&9.64-1&9.64-1
&9.64-1&9.64-1&9.64-1&9.64-1&9.64-1&9.64-1\\
&&9.64-1&9.64-1&9.64-1&9.64-1&9.64-1&9.64-1&9.64-1&9.64-1&9.64-1&9.64-1\\
&&9.64-1&9.64-1&9.64-1&9.64-1&9.64-1&9.64-1&9.64-1&9.64-1&9.64-1&9.64-1\\
&&9.64-1&9.64-1&9.64-1&9.64-1&9.64-1&9.64-1&9.64-1&9.64-1&9.64-1&9.64-1\\
&&9.67-1&9.67-1&9.67-1&9.67-1&\multicolumn{6}{c}{Obs: 1.048$^a$;
1.094$^d$; 8.33-1$^e$; 1.652$^f$}\\
9-4&5177&7.06-1&8.19-1&9.27-1&1.02-0
&7.85-1&8.34-1&8.80-1&8.77-1&9.23-1&9.66-1\\
&&7.04-1&8.18-1&9.26-1&1.02-0&7.84-1&8.34-1&8.80-1&8.76-1&9.23-1&9.66-1\\
&&5.59-1&8.05-1&8.20-1&1.01-0&7.05-1&8.09-1&8.67-1&8.11-1&9.02-1&9.54-1\\
&&1.54-0&1.57-0&1.54-0&1.57-0&1.55-0&1.56-0&1.57-0&1.55-0&1.56-0&1.57-0\\
&&6.12-1&6.33-1&7.77-1&7.99-1&\multicolumn{6}{c}{Obs: 7.39-1$^a$,
6.55-1$^d$,1.478$^f$}\\
7-2&5234&6.55-2&7.23-2&5.65-2&6.10-2
&6.37-2&6.65-2&6.91-2&5.98-2&6.22-2&6.44-2\\
&&6.60-2&7.24-2&5.67-2&6.10-2&6.39-2&6.66-2&6.92-2&5.99-2&6.22-2&6.44-2\\
&&1.25-1&8.51-2&8.39-2&6.53-2&9.30-2&7.94-2&7.80-2&7.97-2&7.05-2&7.00-2\\
&&1.81-1&1.81-1&1.81-1&1.79-1&1.81-1&1.81-1&1.80-1&1.81-1&1.80-1&1.80-1\\
&&&&&&\multicolumn{6}{c}{Obs:}\\
6-1&5278&1.87-1&1.94-1&1.56-1&1.61-1
&1.79-1&1.82-1&1.85-1&1.66-1&1.69-1&1.71-1\\
&&1.92-1&1.95-1&1.58-1&1.61-1&1.81-1&1.83-1&1.85-1&1.67-1&1.69-1&1.71-1\\
&&7.27-1&2.94-1&3.98-1&1.93-1&4.39-1&2.88-1&2.54-1&3.40-1&2.36-1&2.14-1\\
&&5.05-1&4.98-1&5.05-1&4.95-1&5.04-1&5.01-1&4.97-1&5.04-1&5.00-1&4.96-1\\
&&2.19-1&2.21-1&1.83-1&1.84-1&\multicolumn{6}{c}{Obs: 3.20-1$^a$,
5.56-1$^d$}\\
5-1&5335&5.10-1&5.09-1&4.89-1&4.86-1
&5.05-1&5.04-1&5.03-1&4.97-1&4.96-1&4.95-1\\
&&5.66-1&5.14-1&5.06-1&4.88-1&5.23-1&5.10-1&5.07-1&5.08-1&5.00-1&4.97-1\\
&&6.64-0&1.42-0&3.19-0&7.90-1&3.35-0&1.56-0&1.13-0&2.39-0&1.17-0&8.92-1\\
&&1.72-0&1.68-0&1.72-0&1.67-0&1.72-0&1.70-0&1.68-0&1.72-0&1.69-0&1.67-0\\
&&7.17-1&7.09-1&6.25-1&6.20-1&\multicolumn{6}{c}
{Obs: 7.65-1$^a$;1.083$^b$;9.52-1$^c$; 5.67-1$^d$; 1.0$^f$}\\
6-2&5425&4.01-1&4.16-1&3.35-1&3.44-1
&3.84-1&3.90-1&3.96-1&3.57-1&3.62-1&3.66-1\\
&&4.11-1&4.17-1&3.38-1&3.44-1&3.88-1&3.92-1&3.97-1&3.59-1&3.62-1&3.67-1\\
&&1.56-0&6.31-1&8.53-1&4.14-1&9.40-1&6.18-1&5.44-1&7.28-1&5.06-1&4.59-1\\
&&1.08-0&1.07-0&1.08-0&1.06-0&1.08-0&1.07-0&1.07-0&1.08-0&1.07-0&1.06-0\\
&&4.47-1&4.50-1&3.73-1&3.75-1&\multicolumn{6}{c}{Obs: 4.85-1$^a$;
4.33-1$^d$; 7.61-1$^f$}\\
7-3&5427&2.23-1&2.46-1&1.92-1&2.08-1
&2.17-1&2.27-1&2.35-1&2.04-1&2.12-1&2.19-1\\
&&2.25-1&2.46-1&1.93-1&2.08-1&2.18-1&2.27-1&2.36-1&2.04-1&2.12-1&2.19-1\\
&&4.24-1&2.90-1&2.86-1&2.22-1&3.17-1&2.71-1&2.66-1&2.71-1&2.40-1&2.39-1\\
&&6.16-1&6.15-1&6.16-1&6.11-1&6.16-1&6.15-1&6.14-1&6.15-1&6.14-1&6.13-1\\
&&1.57-1&1.62-1&1.39-1&1.43-1&\multicolumn{6}{c}{Obs: 4.34-1$^a$;
3.98-1$^d$}\\
5-2&5485&2.84-1&2.83-1&2.72-1&2.71-1
&2.81-1&2.81-1&2.80-1&2.77-1&2.76-1&2.76-1\\
&&3.15-1&2.86-1&2.82-1&2.72-1&2.91-1&2.84-1&2.82-1&2.83-1&2.78-1&2.77-1\\
&&3.69-0&7.92-1&1.77-0&4.40-1&1.87-0&8.68-1&6.31-1&1.33-0&6.49-1&4.97-1\\
&&9.58-1&9.34-1&9.58-1&9.31-1&9.55-1&9.44-1&9.33-1&9.55-1&9.43-1&9.32-1\\
&&4.00-1&3.93-1&3.47-1&3.44-1&\multicolumn{6}{c}{Obs: 4.60-1$^a$;
6.08-1$^d$; 7.83-1$^f$}\\
6-3&5631&4.21-1&4.37-1&3.51-1&3.61-1
&4.04-1&4.10-1&4.16-1&3.75-1&3.80-1&3.85-1\\
&&4.32-1&4.38-1&3.55-1&3.62-1&4.07-1&4.11-1&4.17-1&3.77-1&3.81-1&3.85-1\\
&&1.64-0&6.62-1&8.96-1&4.35-1&9.87-1&6.49-1&5.71-1&7.64-1&5.31-1&4.82-1\\
&&1.14-0&1.12-0&1.14-0&1.12-0&1.14-0&1.13-0&1.12-0&1.13-0&1.13-0&1.12-0\\
&&4.72-1&4.76-1&3.94-1&3.97-1&\multicolumn{6}{c}{Obs: 4.85-1$^a$;
7.42-1$^e$; 4.78-1$^f$}\\
7-4&5677&4.77-1&5.27-1&4.12-1&4.45-1
&4.64-1&4.85-1&5.04-1&4.36-1&4.54-1&4.70-1\\
&&4.81-1&5.28-1&4.13-1&4.45-1&4.66-1&4.86-1&5.04-1&4.37-1&4.54-1&4.70-1\\
&&9.08-1&6.21-1&6.12-1&4.76-1&6.78-1&5.79-1&5.69-1&5.81-1&5.14-1&5.11-1\\
&&1.32-0&1.32-0&1.32-0&1.31-0&1.32-0&1.32-0&1.32-0&1.32-0&1.32-0&1.31-0\\
&&3.30-1&3.42-1&2.93-1&3.02-1&\multicolumn{6}{c}{Obs: 4.49-1$^a$;
6.67-1$^d$; 3.91-1$^f$}\\
\hline
\end{tabular}
\end{table}

\newpage
\begin{table}
\caption{Line intensity ratios for transitions with common upper level.
A-ratios - ratios of transition probabilities from the present calculation;
NS -  line ratios from Nussbaumer and Storey (1978);
Present -  line ratios from the present results;
Obs - observational line ratios for various planetary nebulae.}
\begin{tabular} {cccccc}
\hline
Level Index&Wavelengths&A-ratios&NS&CAL&OBS\\
\hline
$\frac{5-1}{5-2}$&$\frac{I(5335)}{I(5485)}$&1.748&1.793&1.797&1.663$^a$,
0.933$^d$,1.277$^f$\\
\hline
$\frac{6-1}{6-2}$&$\frac{I(5278)}{I(5425)}$&0.454&0.490&0.467&0.660$^a$,
1.284$^d$\\
\hline
$\frac{6-2}{6-3}$&$\frac{I(5425)}{I(5631)}$&0.917&0.947&0.952&1.0$^a$,
1.592$^f$\\
\hline
$\frac{7-2}{7-3}$&$\frac{I(5234)}{I(5427)}$&0.283&&0.294&\\
\hline
$\frac{7-3}{7-4}$&$\frac{I(5427)}{I(5677)}$&0.446&0.474&0.467&0.967$^a$,
0.587$^d$\\
\hline
$\frac{8-2}{8-3}$&$\frac{I(4973)}{I(5146)}$&0.932&0.967&0.964&1.048$^a$,
1.094$^d$,0.833$^e$,1.652$^f$\\
\hline
\end{tabular}
\\
$^a$ Hyung and Aller (1997a);
$^c$ Hyung \etal (1997);
$^d$ Feibelman \etal (1996);\\
$^e$ Hyung and Aller (1998);
$^f$ Hyung and Aller (1997b).
\end{table}
\clearpage

\newpage
\begin{figure}
\centering
\psfig{figure=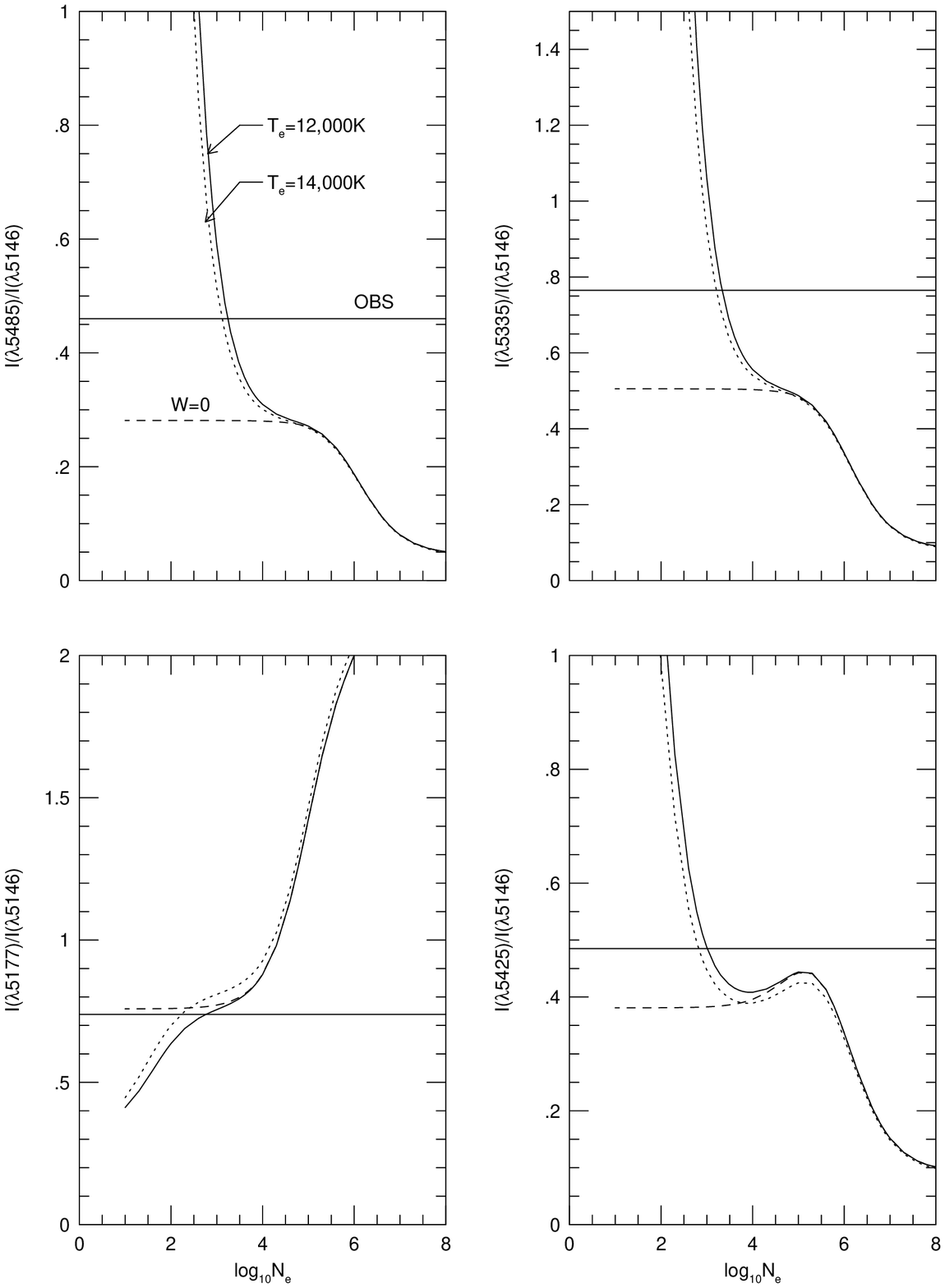,height=20.0cm,width=18.0cm}
\caption{NGC 6741: Line ratios with fluorescent excitation (FLE), with
T$_{eff}$ = 140,000 K, W = 10$^{-14}$, at T$_e$ = 12,000
and 14,000 K -- solid and dotted 
lines respectively; without FLE (T$_e$ =
12,000 K), W = 0 - dashed line; Observed values from sources in the
text (Table~5) - OBS.}
\end{figure}

\pagebreak
\begin{figure}
\centering
\psfig{figure=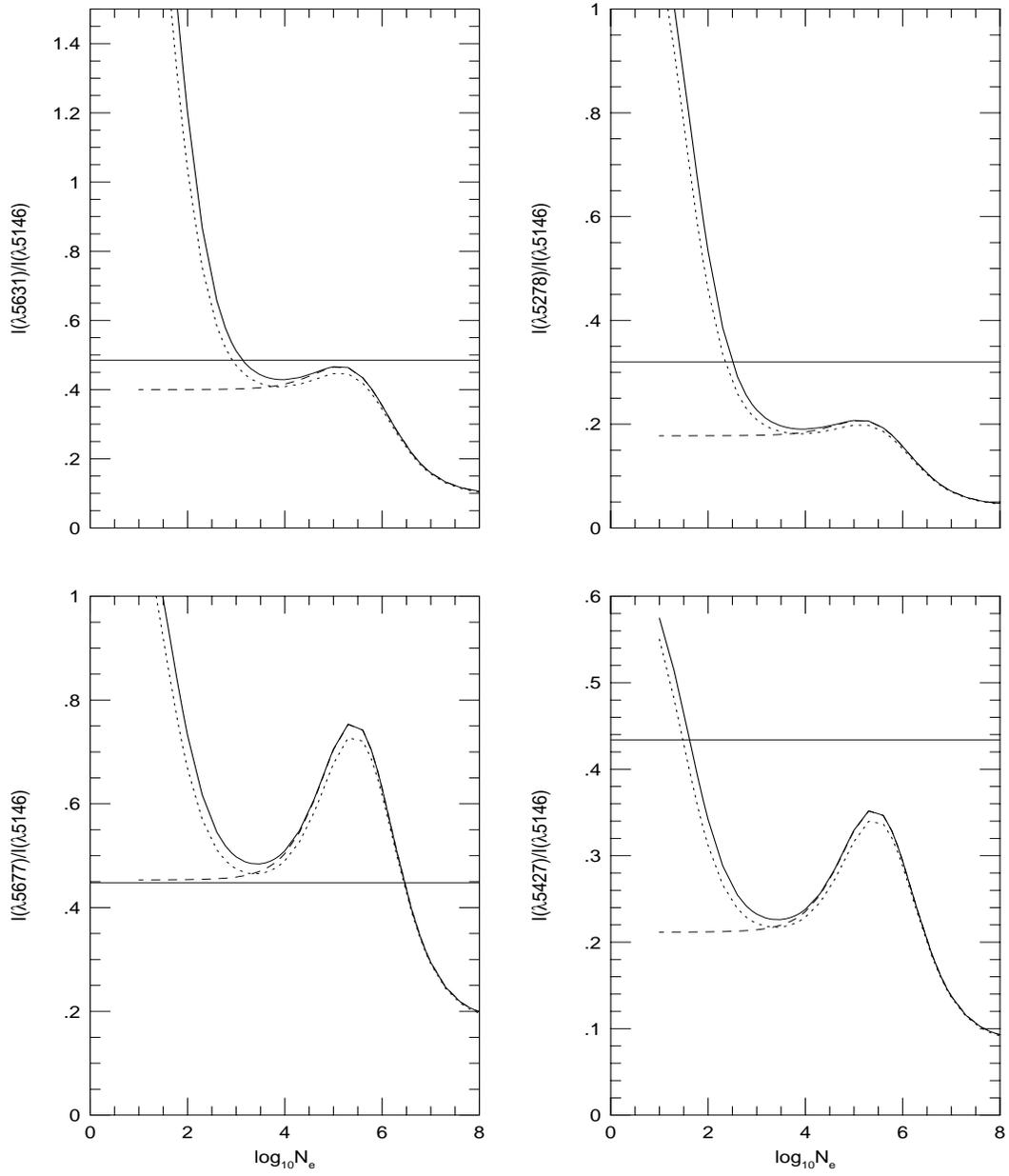,height=20.0cm,width=18.0cm}
\caption{NGC 6741: Line ratios with and without FLE, as in Fig. 1.}
\end{figure}

\pagebreak
\begin{figure}
\centering
\psfig{figure=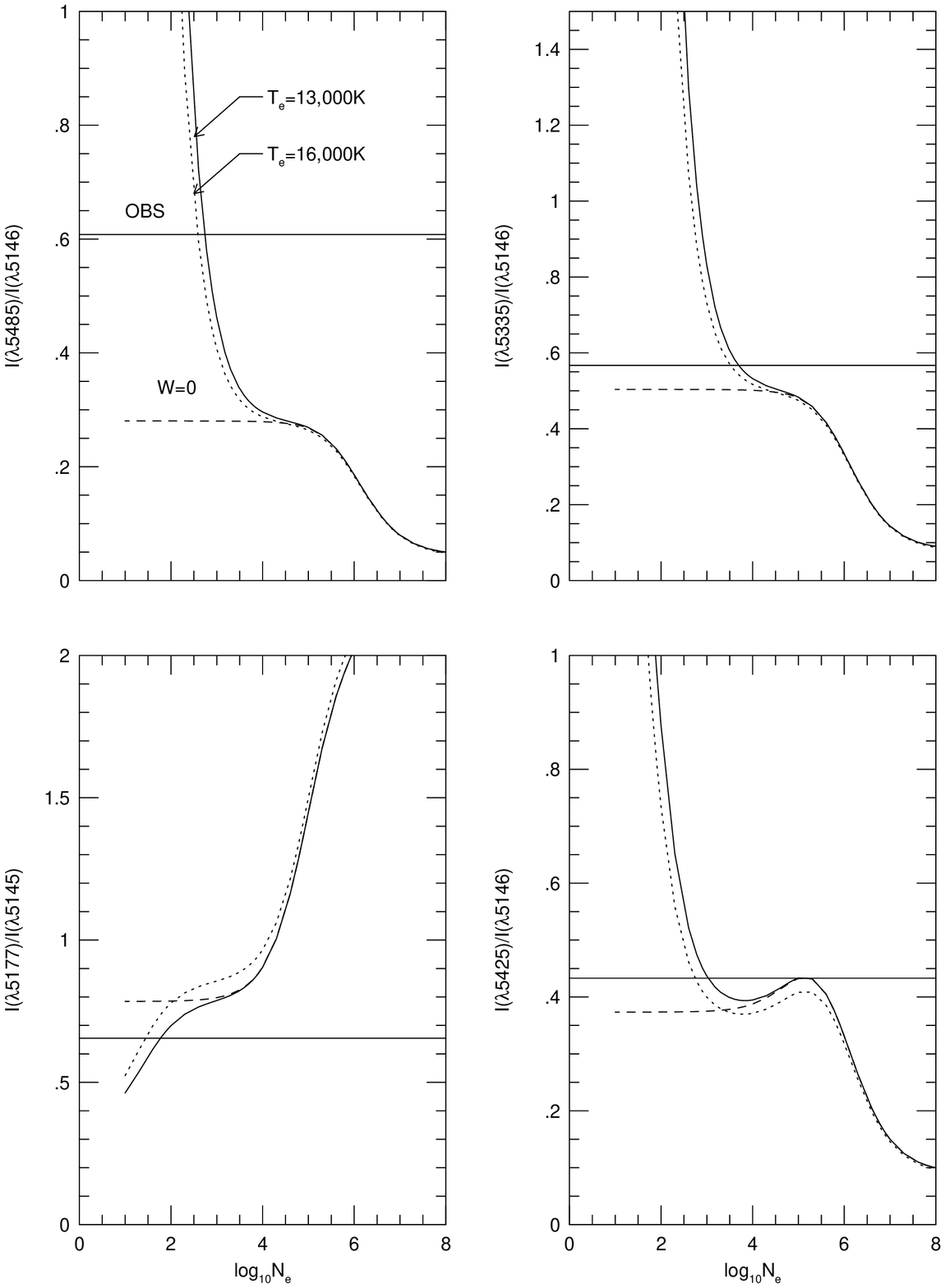,height=20.0cm,width=18.0cm}
\caption{IC 351: Line ratios with fluorescent excitation (FLE), with
 T$_{eff}$ = 105,000 K, W = 10$^{-14}$, at T$_e$ =
13,000 and 16,000 K -- solid and 
dotted lines respectively; without FLE (T$_e$ =
13,000 K), W = 0 - dashed line; Observed values from sources in the text
(Table~5) - OBS.}
\end{figure}

\pagebreak
\begin{figure}
\centering
\psfig{figure=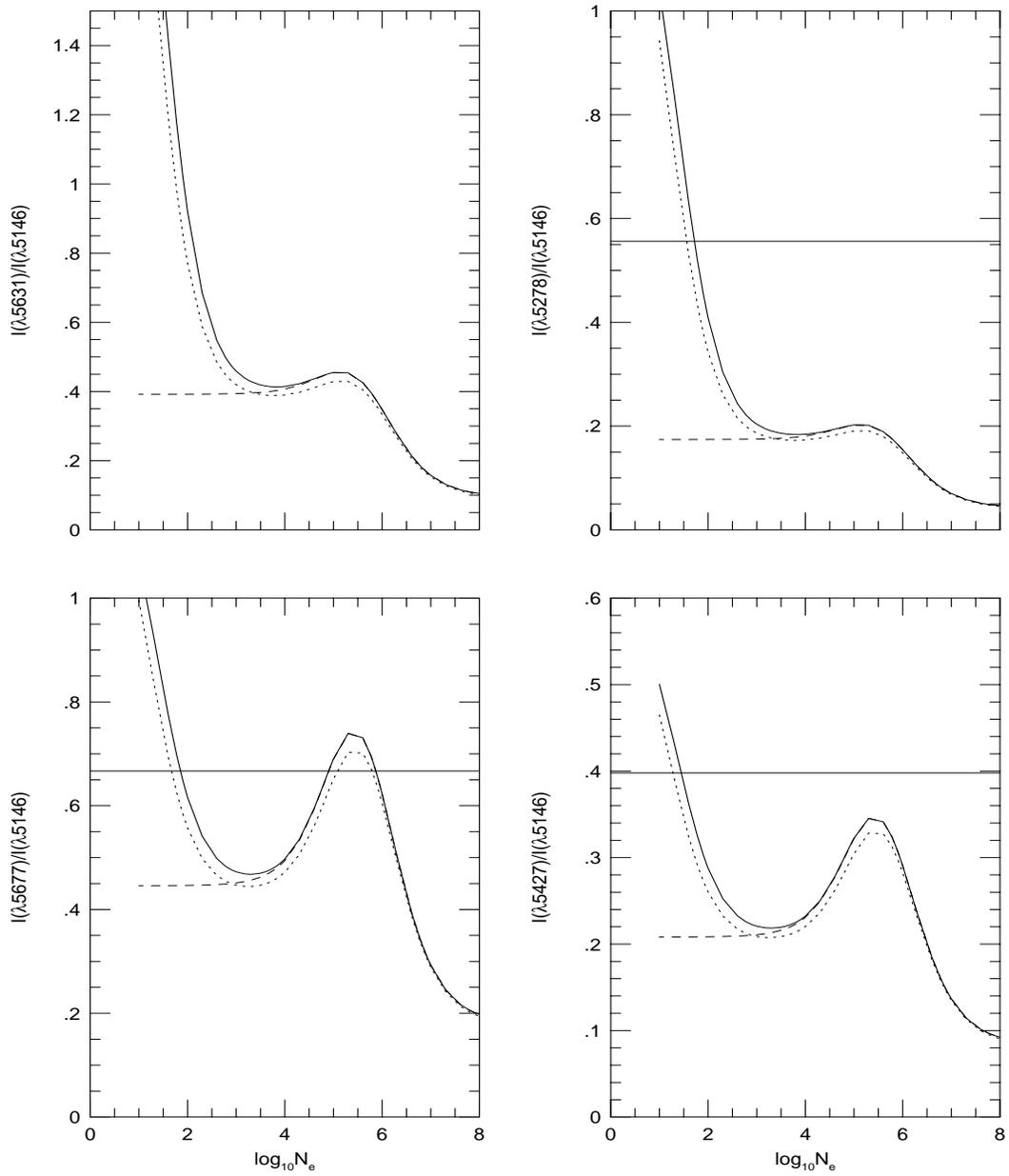,height=20.0cm,width=18.0cm}
\caption{IC 351: Line ratios with and without FLE, as in Fig. 3.}
\end{figure}

\pagebreak
\begin{figure}
\centering
\psfig{figure=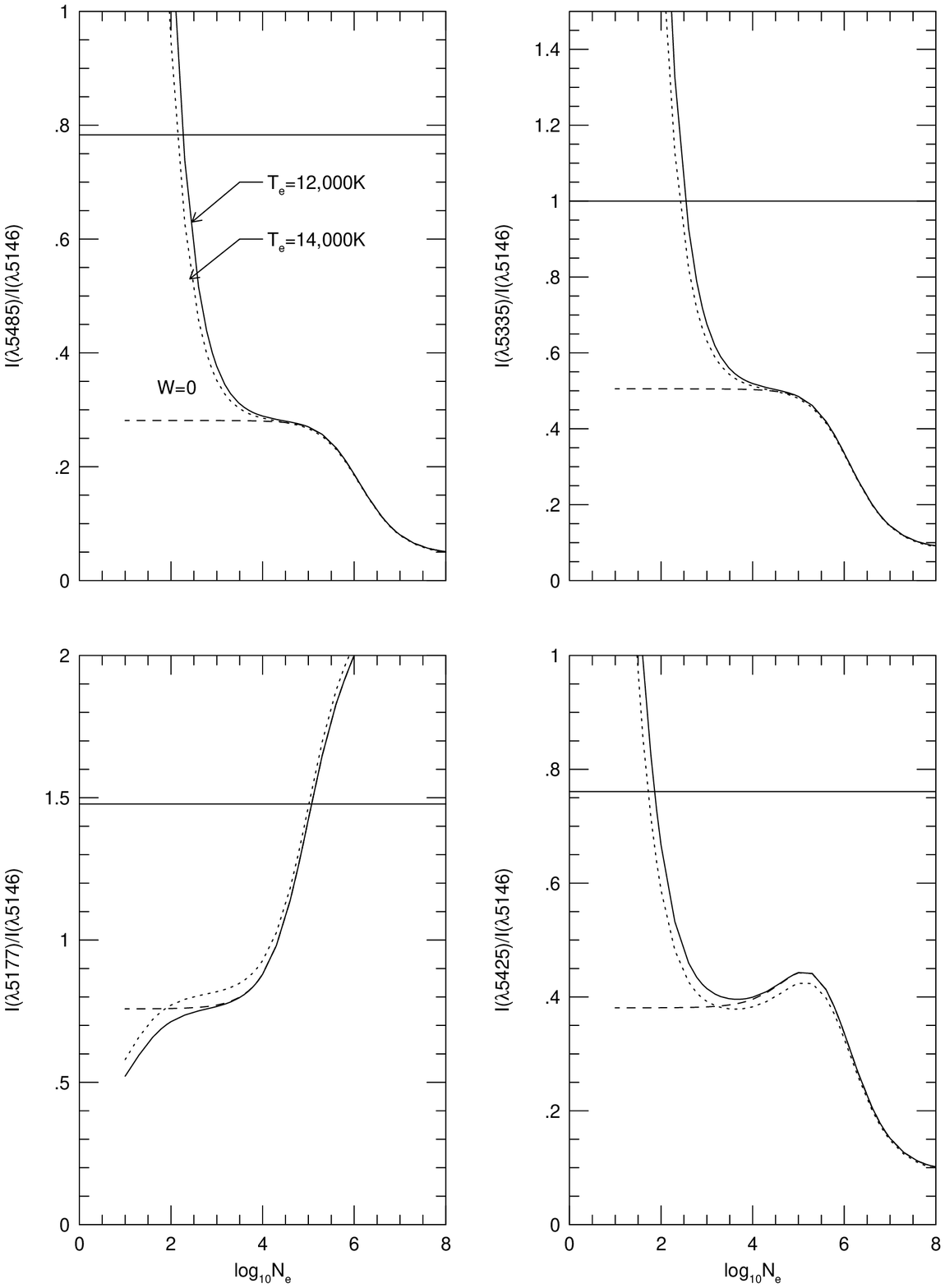,height=20.0cm,width=18.0cm}
\caption{NGC 7662: Line ratios with fluorescent excitation (FLE), with
T$_{eff}$ = 80,000 K, W = 10$^{-13}$, at T$_e$ =
12,000 and 14,000 K -- solid and 
dotted lines respectively; without FLE (T =
12,000 K), W = 0 - dashed line; Observed values from sources in the text
(Table~5) - OBS.}
\end{figure}

\pagebreak
\begin{figure}
\centering
\psfig{figure=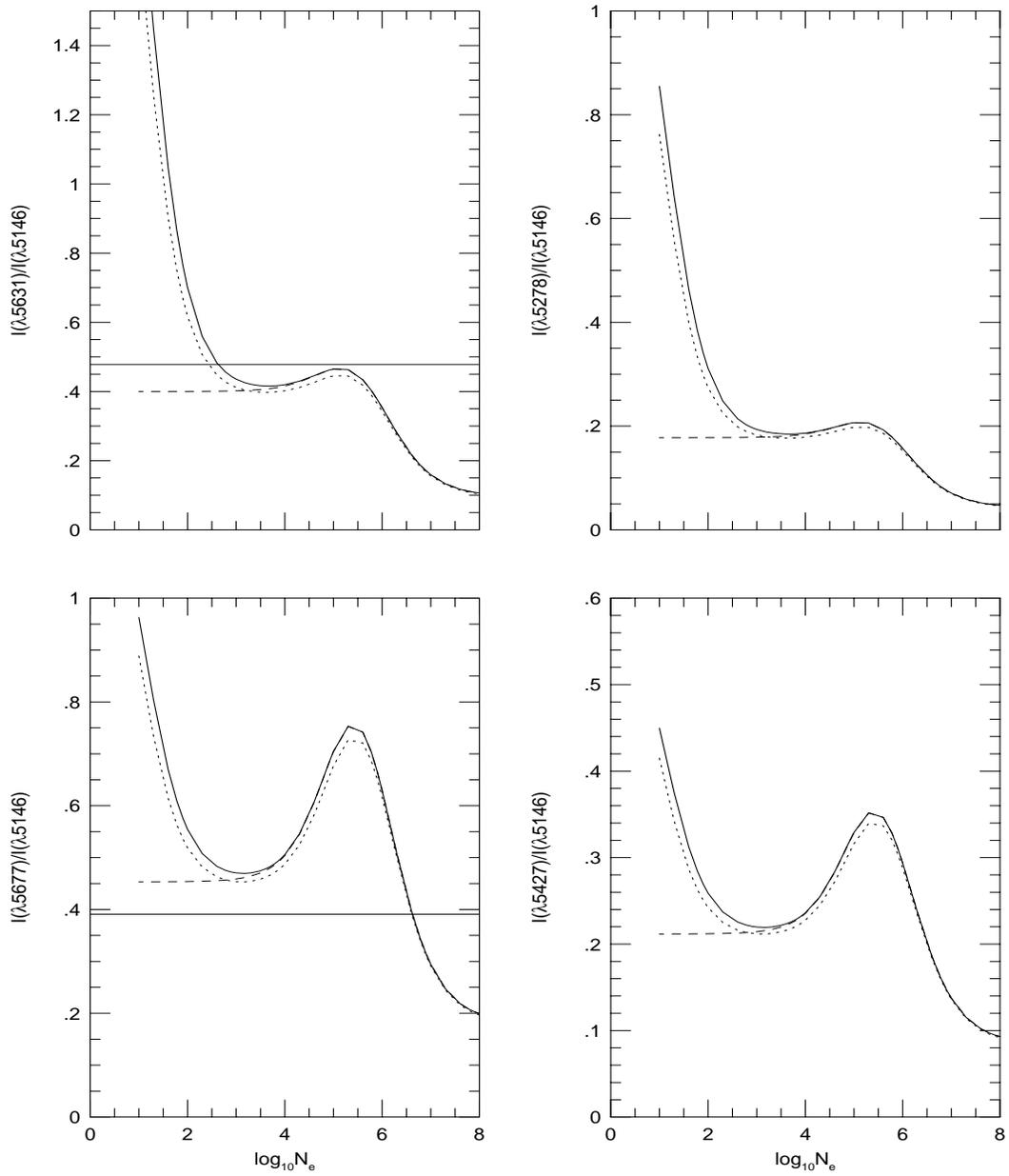,height=20.0cm,width=18.0cm}
\caption{NGC 7662: Line ratios with and without FLE, as in Fig. 5.}
\end{figure}

\pagebreak
\begin{figure}
\centering
\psfig{figure=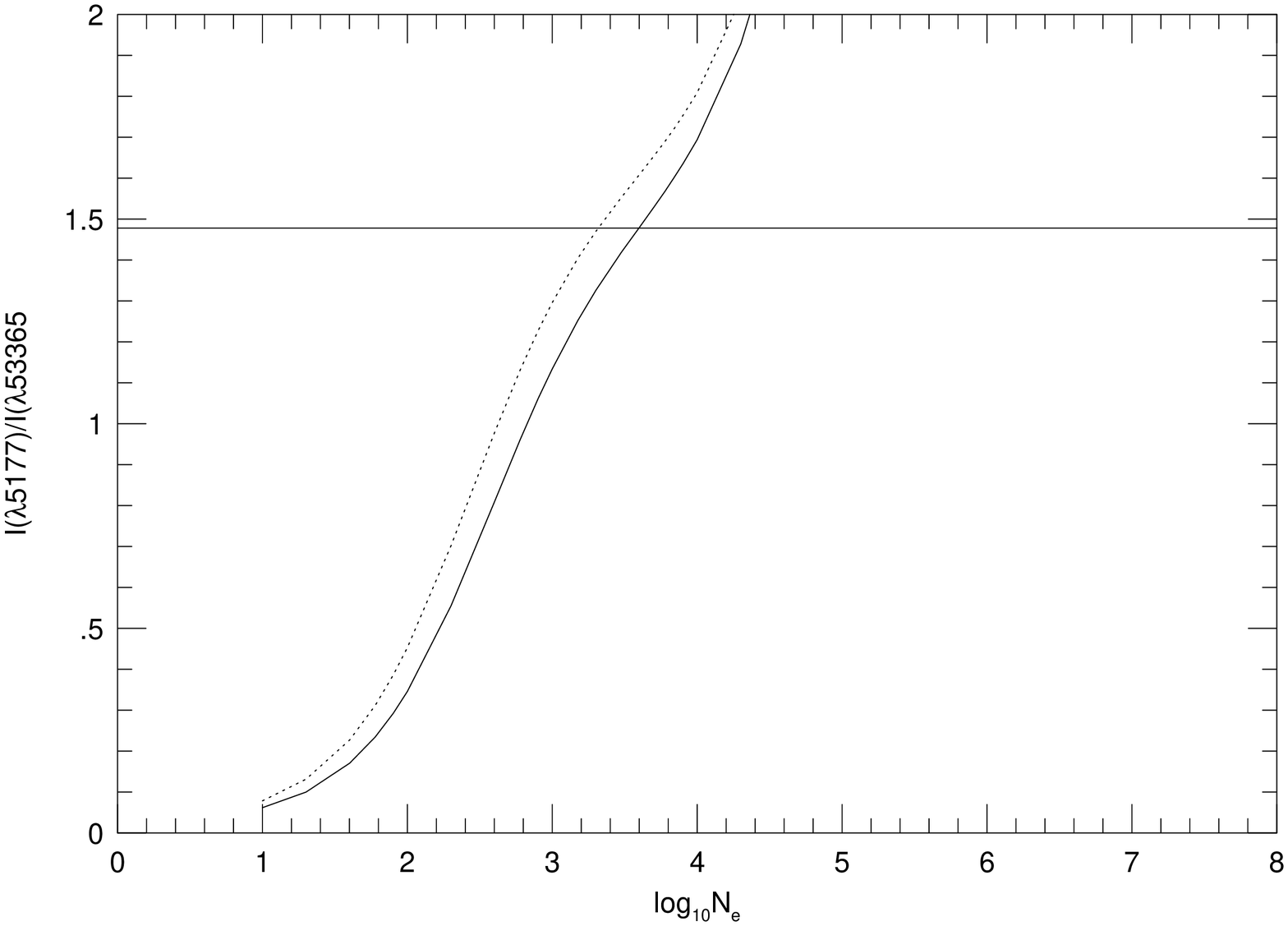,height=10.0cm,width=18.0cm}
\caption{NGC 7662: Line ratio 5177/5335 with FLE, as in Figs. 5
and 6.}
\end{figure}

\pagebreak
\begin{figure}
\centering
\psfig{figure=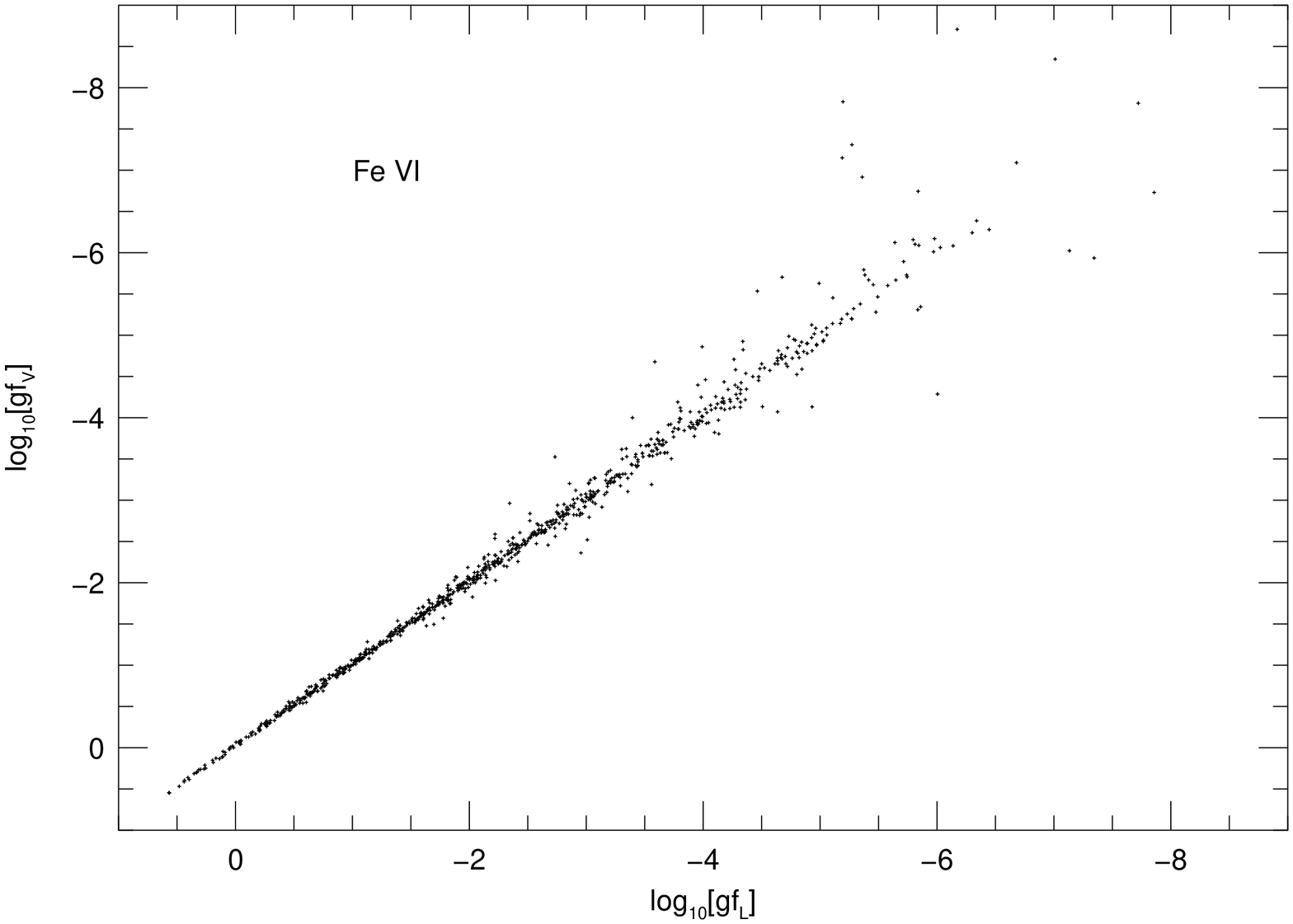,height=10.0cm,width=18.0cm}
\caption{Comparison of Length vs. Velocity gf-values for 867 dipole allowed
and intercombination $E1$ transitions in Fe~VI.}
\end{figure}

\end{document}